\documentclass[runningheads]{llncs}
\usepackage[T1]{fontenc}
\usepackage{graphicx}
\usepackage{etoolbox}

\usepackage{hyperref} 
\usepackage[firstpage]{draftwatermark} 
\SetWatermarkAngle{0}

\SetWatermarkText{\raisebox{14.7cm}{%
		\hspace{0.1cm}%
		\href{https://zenodo.org/records/10927672}{\includegraphics{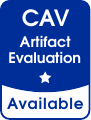}}%
		\hspace{9cm}%
		\includegraphics{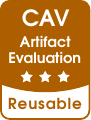}%
}}

\newtoggle{arxiv}
\newcommand{\ifarxivelse}[2]{\iftoggle{arxiv}{#1}{#2}}
\toggletrue{arxiv} 

\usepackage{amsmath,amsfonts,amssymb}
\usepackage{booktabs}
\usepackage{tikz}
\usepackage{csquotes}
\usepackage{thmtools}
\usepackage{thm-restate}

\usepackage{pgfplots}

\usepackage{tabu,multirow}
\usepackage{xparse}
\usepackage{mathtools}
\usepackage{xfrac}

\usepackage{algorithmicx,algorithm}
\usepackage[noend]{algpseudocode}

\usepackage[capitalize]{cleveref}

\usepackage{subcaption}
\usepackage{paralist}

\usepackage{placeins} 

\newcommand{\intersectionSym}{\cap}
\newcommand{\intersectionBin}{\mathbin{\intersectionSym}}
\newcommand{\UnionSym}{\bigcup}

\newcommand{\intersection}{\intersectionBin}
\newcommand{\Union}{\UnionSym}

\DeclareGraphicsExtensions{.pdf, .png, .jpg}

\newcommand{\expec}{\mathbb{E}}

\newcommand{\Naturals}{\mathbb{N}}
\newcommand{\Reals}{\mathbb{R}}

\newcommand{\eqdef}{\vcentcolon=}

\newcommand{\MC}{\mathsf{M}}

\newcommand{\infinitepath}{\rho}
\NewDocumentCommand{\Infinitepaths}{d<>}{\IfNoValueTF{#1}{\mathsf{Paths}}{\mathsf{Paths}_{#1}}}

\newcommand{\distribution}{d}
\NewDocumentCommand{\Distributions}{d()}{\IfNoValueTF{#1}{\mathcal{D}}{\mathcal{D}(#1)}}

\newcommand{\maxsymbol}{\bigtriangleup}
\newcommand{\minsymbol}{\bigtriangledown} 

\newcommand{\G}{\mathsf{G}}
\newcommand{\allstates}{\mathsf{S}}
\newcommand{\states}{\allstates}
\newcommand{\maxstates}{\allstates_{\maxsymbol}}
\newcommand{\minstates}{\allstates_{\minsymbol}}
\newcommand{\initstate}{\hat{s}}
\newcommand{\act}{\mathsf{A}}

\newcommand{\state}{s}
\newcommand{\trans}{\delta}

\newcommand{\statereward}{\mathsf{r}}

\newcommand{\straa}{\sigma}
\newcommand{\strab}{\tau}

\newcommand{\probability}{\mathbb{P}}

\newcommand{\val}{\mathsf{V}}
\newcommand{\lb}{\mathsf{L}}
\newcommand{\ub}{\mathsf{U}}
\newcommand{\bellman}{\mathcal{B}}

\newcommand{\opt}{\mathsf{opt}}

\newcommand{\tool}[1]{\textsc{#1}\xspace}
\newcommand*{\pet}{\tool{Pet}}
\newcommand*{\petone}{\tool{Pet1}}
\newcommand*{\pettwo}{\tool{Pet2}}
\newcommand*{\prism}{\tool{Prism}}
\newcommand*{\prismgames}{\tool{Prism-games}}
\newcommand*{\prismext}{\tool{Prism-ext}}
\newcommand*{\storm}{\tool{Storm}}
\newcommand*{\tempest}{\tool{Tempest}}
\newcommand{\model}[1]{\texttt{#1}\xspace}

\usetikzlibrary{shapes,positioning,arrows,arrows.meta,calc,automata,matrix,fit,backgrounds,fadings,colorbrewer,plotmarks}

\usepackage{tikzsymbols}
\pgfplotsset{compat=1.17}

\tikzstyle{loopaction}=[font=\small,outer sep=2pt]
\tikzstyle{actionnode}=[circle,draw=black,fill=black,minimum size=1mm,inner sep=0,outer sep=0]
\tikzstyle{actionedge}=[draw,-]
\tikzstyle{prob}=[font=\scriptsize,outer sep=1pt]
\tikzstyle{probedge}=[draw,->]
\tikzstyle{directedge}=[draw,->]

\tikzset{chainarrow/.tip={Stealth[length=3pt]}}
\tikzset{>=chainarrow}

\tikzset{
	action/.style={font=\small,outer sep=2pt,inner sep=0pt},
	state/.style={thick,minimum width=0.75cm,minimum height=0.75cm,inner sep=0.1em,rectangle,rounded corners,black,draw},
	triangle/.style={regular polygon, regular polygon sides=3},
	max vertex/.style={state,path picture={\draw[gray,rounded corners=0,fill,fill opacity=0.2,thick] (path picture bounding box.south west) -- (path picture bounding box.north) -- (path picture bounding box.south east);}},
	min vertex/.style={state,path picture={\draw[gray,rounded corners=0,fill,fill opacity=0.2,thick] (path picture bounding box.north west) -- (path picture bounding box.south) -- (path picture bounding box.north east);}},
	reward/.style={inner sep=0pt,outer sep=0.2ex},
}

\newcommand{\stateandreward}[2]{{#1}{:}{#2}}
\crefname{section}{Sec.}{Secs.}%
\crefname{appendix}{App.}{Apps.}%
\crefname{lemma}{Lem.}{Lemms.}
\crefname{theorem}{Thm.}{Thms.}
\crefname{corollary}{Cor.}{Cors.}
\crefname{equation}{Eq.}{Eqs.}
\crefname{figure}{Fig.}{Figs.}
\crefname{tabular}{Tab.}{Tabs.}

\Crefname{section}{Sec.}{Secs.}%
\Crefname{appendix}{App.}{Apps.}%
\Crefname{lemma}{Lem.}{Lemms.}
\Crefname{theorem}{Thm.}{Thms.}
\Crefname{corollary}{Cor.}{Cors.}
\Crefname{equation}{Eq.}{Eqs.}
\Crefname{figure}{Fig.}{Figs.}
\Crefname{tabular}{Tab.}{Tabs.}

\usepackage{todonotes}

\begin{document}
\title{Playing Games with your PET: Extending the Partial Exploration Tool to Stochastic Games\thanks{M.\ Weininger has received funding from the EU's Horizon 2020 research and innovation programme under the Marie Skłodowska-Curie grant agreement No.\ 101034413.
\\
\textbf{Data availability:} We refer to the artefact with the exact code used for the submission and all logs~\cite{artifact}, and the gitlab with the continually developed source code~\cite{pet-gitlab}.
}}
\titlerunning{Extending PET for Solving SGs}
%
\author{Tobias Meggendorfer\inst{1}\orcidID{0000-0002-1712-2165}\\ \and
Maximilian Weininger\inst{2}\orcidID{0000-0002-1825-0097}}
\authorrunning{T.~Meggendorfer and M.~Weininger}
%
\institute{
Lancaster University Leipzig, Leipzig, Germany\\
\email{tobias@meggendorfer.de}
\and
Institute of Science and Technology Austria, Klosterneuburg, Austria\\
\email{mweining@ista.ac.at}}
\maketitle              
\begin{abstract}
We present version 2.0 of the \emph{Partial Exploration Tool} (\pet), a tool for verification of probabilistic systems.
We extend the previous version by adding support for \emph{stochastic games}, based on a recent unified framework for sound value iteration algorithms.
Thereby, \pettwo is the first tool implementing a sound and efficient approach for solving stochastic games with objectives of the type reachability/safety and mean payoff.
We complement this approach by developing and implementing a partial-exploration based variant for all three objectives.
Our experimental evaluation shows that \pettwo offers the most efficient partial-exploration based algorithm and is the most viable tool on SGs, even outperforming unsound tools.

\keywords{Probabilistic verification \and Stochastic games \and Partial exploration \and Model checker}
\end{abstract}
\section{Introduction}\label{sec:intro}
\noindent Stochastic games (SGs)~\cite{condonAlgo} are a foundational model for sequential decision making in the presence of uncertainty and two antagonistic agents.
They are practically relevant, with applications ranging from economics~\cite{amir2003stochastic} over IT security~\cite{DBLP:conf/hicss/RoyESDSW10} to medicine~\cite{bellomo2008mathematical}; and they are theoretically fundamental, in particular because many associated classical decision problems are representative of the important complexity class NP~$\cap$~co-NP, e.g.\ deciding whether the value of a reachability or mean payoff objective is greater than a given threshold~\cite{DBLP:journals/talg/Johnson07,DBLP:conf/isaac/AnderssonM09}
 (see \cite{DBLP:conf/soda/ChatterjeeMSS23} for recent advances).
See~\cite[Chp.~1.3]{MaxiThesis} for further motivation.

However, even mature tools either do not support SGs at all (\storm{}~\cite{storm}) or employ approaches without formal guarantees, i.e.\ their results can be wrong (\prismgames~\cite{prism-games} and \tempest~\cite{tempest}), which is unacceptable in the context of safety-critical applications.
%
This is because \emph{value iteration} (VI), the de-facto standard approach to solving stochastic systems, lacks a sound and efficient \emph{stopping criterion}, i.e.\ a \enquote{rule} to check whether the current iterates are sufficiently close to the correct value.
For Markov decision processes (MDPs) (SGs with only one player) such a sound variant of VI (often called \emph{interval iteration}) was developed a decade ago~\cite{atva,HM18} and subsequently implemented in practically all major model checkers. 
However, extending the underlying reasoning to SGs proved to be surprisingly tricky, with sound variants even for special cases only developed quite recently~\cite{EKKW22}. 
Just a year ago, \cite{lics23} presented a unified way of ensuring the soundness of VI for solving SGs with various quantitative objectives, which forms the theoretical basis for this work.%

Note that the classical approaches strategy iteration and quadratic programming are sound in theory, but (i)~the available implementations of~\cite{gandalf} are prototypical and unsound \cite[Sec.~5]{gandalf}, and (ii)~these approaches usually either are not practically efficient or use heuristics that actually make them unsound
~\cite{DBLP:conf/tacas/HartmannsJQW23}.

\paragraph{Contributions.}
We present version 2.0 of the \emph{Partial Exploration Tool} (\pet), the first tool implementing a sound and efficient approach for solving SGs with objectives of the type reachability/safety and mean payoff (a.k.a.\ long-run average reward).
In the following, we write \petone and \pettwo to refer to the previous and now presented version of \pet, respectively.

\emph{Theoretically}, \pettwo is based on the results of \cite{lics23}.
We provide two flavours of their approach:
Firstly, we implement the basic complete-exploration (CE) algorithm, enhanced with several theoretical improvements, both new and suggested in the literature.
Secondly, we develop a \emph{partial-exploration (PE)} approach, the focus of \pettwo, by combining the ideas of~\cite{lics23} with those in~\cite{atva,EKKW22,DBLP:conf/atva/Meggendorfer22}.

\emph{Practically}, \pettwo is an extension of \petone~\cite{DBLP:conf/atva/Meggendorfer22} (only applicable to MDPs).
Apart from adding support for dealing with SGs and completely replacing the approach of \petone with the ideas of \cite{lics23}, we implemented many engineering improvements.
Concretely, despite employing a more general algorithm, our experimental evaluation shows that \pettwo's performance is on par with \petone.
Moreover, \pettwo outperforms the existing SG solvers \prismgames and \tempest, despite those not providing guarantees (and indeed returning wrong results).

\section{Preliminaries}\label{sec:prelims}
%
%
\noindent Here, we very briefly recall turn-based stochastic games as far as they are necessary to understand this paper, with more details in \ifarxivelse{\Cref{app:prelims}}{\cite[App.~A]{pet2-techreport}} and \cite{lics23}.

A \emph{(turn-based) stochastic game (SG)} (e.g.\ \cite{condonAlgo}) consists of a set of states $\states$ that belong to either the \emph{Maximizer} 
or \emph{Minimizer} player; 
a set of available actions for every state, denoted $\act(s)$; 
and a probabilistic transition function $\trans$ that for a state-action pair gives a probability distribution over successor states.
An SG where all states belong to one player is called \emph{Markov decision process (MDP)}, see~\cite{DBLP:books/wi/Puterman94}; without nondeterministic choices, it is a \emph{Markov chain (MC)}.

SGs are played in turns as follows:
Starting in an \emph{initial state} $s_0$, the player to whom this state belongs chooses an action $a_0 \in \act(s)$. Then, the play advances to the next state $s_1$, which is sampled according to the probability distribution given by $\trans(s,a)$.
Repeating this process indefinitely yields an infinite path $\infinitepath = s_0 a_0 s_1 a_1 \dots$.
We write $\Infinitepaths<\G>$ for the set of all such infinite paths in a game $\G$.

A \emph{memoryless deterministic (MD) strategy} $\straa$ of Maximizer assigns an action to every Maximizer state $s$, i.e.\ $\straa(s) \in \act(s)$.
Minimizer strategies $\strab$ are defined analogously.
By fixing a pair of strategies $(\straa,\strab)$ and thereby resolving all non-deterministic choices, we obtain a Markov chain that together with an initial state $\initstate$ induces a unique probability distribution over the set of all infinite paths $\Infinitepaths<\G>$ \cite[Sec.~10.1]{DBLP:books/daglib/0020348}. 
For a random variable over paths $\Phi : \Infinitepaths<\G> \to \Reals$ we write $\expec_{\G,\initstate}^{\straa,\strab}[\Phi]$ for its expected value under this probability measure.


An \emph{objective}  $\Phi : \Infinitepaths<\G> \to \Reals$ formalizes the \enquote{goal} of both players by assigning a value to each path.
In this paper, we focus on \emph{mean payoff} (also called long-run average reward)~\cite{gillette1957stochastic}, which assign to every path the average reward that is obtained in the limit.
The presented algorithms and tools can also explicitly handle reachability/safety objectives, which compute the probability of reaching a given set of states while avoiding another.
Such objectives are special cases of mean payoff, see e.g.~\cite{DBLP:conf/cav/AshokCDKM17}.
Another prominent objective is \emph{total reward}~\cite{CFK+13reward}, but this is practically incompatible with our approach and goals (see \ifarxivelse{\cref{app:TR}}{\cite[App.~B]{pet2-techreport}}).

Given an SG and an objective, we want to compute the \emph{value of the game}, i.e.\ the optimal value the players can ensure by choosing optimal strategies.
Formally, the value of state $s$ is defined as
$
\val_{\G, \Phi}(s) \eqdef {\sup}_\straa~{\inf}_\strab~\expec_{\G, s}^{\straa, \strab}[\Phi] (= {\inf}_\strab~{\sup}_\straa~\expec_{\G, s}^{\straa, \strab}[\Phi]).
$
We are interested in \emph{approximate solutions}, i.e.\ given a concrete state $\hat{s}$ and precision requirement $\varepsilon$, our goal is to determine a number $v$ such that $|\val_{\G, \Phi}(\hat{s}) - v| < \varepsilon$.

An \emph{end component (EC)} intuitively is a set of states in which the system \emph{can} remain forever, given suitable strategies.
Inclusion-maximal ECs are \emph{maximal end components (MECs)}.
The play of an SG eventually remains inside a single MEC with probability one~\cite{dA97}.
In other words, MECs capture all relevant long-run behaviour of the system.
The set of MECs can be identified in PTIME~\cite{CY95}.

\section{Complete-Exploration Algorithm for Solving SGs}\label{sec:CE}

\noindent
In this section, we very briefly recall Alg.~2 of~\cite{lics23}, a generic value-iteration based approach for SGs, which in particular is 
the first such algorithm that provides guarantees on the precision for mean payoff.
This recapitulation is the basis for the following descriptions of both (i)~the main practical improvements over~\cite[Alg. 2]{lics23} added in our implementation (see the end of this section); and (ii)~the new partial-exploration approach described in \Cref{sec:PE}.


\paragraph{Intuition.}
The key insight of \cite{lics23} is to \enquote{split} the analysis of SGs into infinite and transient behaviour.
Infinite behaviour occurs in ECs where both players want to remain under optimal strategies; this is where the mean payoff is actually \enquote{obtained}.
Transient behaviour occurs in states that are not part of such an EC, i.e.\ that are almost surely only visited finitely often; such states in turn achieve their value by trying to reach ECs that give them the best mean payoff.


\paragraph{Algorithm.}
Based on this intuition, we can summarize the overall structure of the algorithm.
It maintains two functions $\lb$ and $\ub$, which map every state to a lower and upper bound on its true value, respectively.
Our aim is to improve these bounds until they are sufficiently close to each other; then we can derive the correct value up to precision $\varepsilon$.
After initializing the bounds to safe under- and over-approximations (e.g.\ the minimum and maximum reward occurring in the SG), we repeatedly perform two operations, further described below:
Firstly, we use so-called \emph{Bellman updates} to back-propagate bounds through the SG, which also is the classical \enquote{value iteration step}.
This corresponds to the transient behaviour, intuitively computing the optimal choice of actions to reach the ECs with the best value.
Secondly, we use the operations of \emph{deflating} and \emph{inflating}. 
These essentially inform states about the value obtainable by staying.

\paragraph{Bellman Updates.}
Bellman updates are at the core of all value iteration style algorithms (see e.g.\ \cite{DBLP:conf/spin/ChatterjeeH08}).
Intuitively, they update the current estimates by \enquote{taking one step}, i.e.\ computing the expectation of following the action that is optimal according to the current estimates.
Formally, given a function $x \colon \states \to \Reals$, the Bellman update $\bellman$ computes a new estimate function as $\bellman(x)(s) \eqdef \opt^s_{a \in \act(s)} \sum_{s'\in\states}\trans(s,a)(s') \cdot x(s')$, where $\opt^s = \max$ if $s$ is a Maximizer state and $\min$ otherwise.
Importantly, if $x$ is a correct lower or upper bound on the true value, then $\bellman(x)(s)$ is, too.
However, only applying $\bellman(x)$ may not converge!

\begin{figure}[t]
	\centering
	\begin{tikzpicture}
		\node[max vertex] at (0,0) (s) {$\stateandreward{s}{4}$};
		\node[cloud,draw=gray,fill=gray!10,cloud puffs=15,aspect=1.75,inner sep=0pt] at (2,0) (t) {$\ub = 5$};

		\path[->,directedge]
			(s) edge[loop left] node[anchor=east,action] {$a$}
			(s) edge node[anchor=south,action] {$b$} (t)
		;
	\end{tikzpicture}\hspace{2cm}
	\begin{tikzpicture}
		\node[cloud,draw=gray,fill=gray!10,cloud puffs=15,aspect=1.5] at (0,0) (q) {$X$};
		\node[min vertex] at (1.5,0) (p) {$p$};
		\node[max vertex] at (3,0) (s) {$s$};
		\node[cloud,draw=gray,fill=gray!10,cloud puffs=15,aspect=1.5] at (4.5,0) (t) {$Y$};

		\path[->,directedge]
			(p) edge node[anchor=south,action] {$x$} (q)
			(p) edge[bend left=15] node[anchor=south,action] {$y$} (s)
			(s) edge[bend left=15] node[anchor=north,action] {$a$} (p)
			(s) edge node[anchor=south,action] {$b$} (t);
	\end{tikzpicture}
	\caption{Example SGs to explain deflating and inflating.
	States with upward and downward triangles denote Maximizer and Minimizer states, respectively.}
	\label{fig:3-example}
\end{figure}
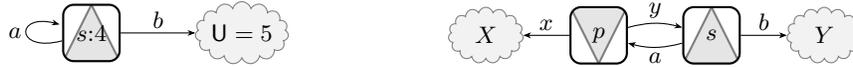

\paragraph{Deflating and Inflating.}
We briefly describe why the convergence problem arises and how deflating and inflating solve the issue, summarizing~\cite[Sec. V-C]{lics23}.


Recall the intuition that a state can get its value either from infinite behaviour (i.e.\ \enquote{staying} in the current area of the game) or from transient behaviour (\enquote{leaving} the current area).
As a simple example, consider the SG (even MDP) in \Cref{fig:3-example} (left).
Assume we have $\ub = 5$ in the grey area, i.e.\ by taking action $b$ we obtain at most $5$, but for $s$ the current upper bound is the conservative over-approximation $\ub(s)=10$.
The Bellman update prefers $a$ over $b$ and keeps $\ub(s)$ at $10$, even though following $a$ forever will only yield a mean payoff of $4$.
This is due to a cyclic dependency: $s$ \enquote{believes} it can (eventually) achieve $10$ because it \enquote{promises} to achieve $10$ by having $\ub(s) = 10$.
Deflating now identifies that Maximizer wants to \enquote{stay} in $s$ using $a$, computes the actual value that is obtained by staying, i.e.\ $4$, and compares it to the best possible exit from this region, namely leaving with $b$ to obtain at most $5$.
Maximizer has to either stay forever or eventually leave, thus we can decrease the upper bound to the maximum of these two actions, i.e.\ $5$.
Dually, inflating raises the value of Minimizer states to the minimum of leaving and staying.
In other words, de-/inflating complement Bellman updates by informing players about the consequences of staying forever, forcing them to choose between this staying value and the best exit.

While this reasoning is simple for single-state, single-player cycles, it gets much more involved in the general case of stochastic games.
In particular, and in contrast to MDPs, in two-player SGs the opponent can restrict which cycles and exits are reachable from certain states.
For example, consider the SG in \Cref{fig:3-example} (right):
States $p$ and $s$ can form a cycle.
However, if $s$ goes to $p$, it depends on the choice of Minimizer in $p$ whether the play stays in the cycle $\{p,s\}$ or leaves towards $X$.
To tackle this issue, \cite[Alg.\ 2]{lics23} repeatedly identifies regions where players want to remain \emph{based on their current estimates}, called \emph{simple end component (SEC)}-candidates.
It does so by fixing one player's choices (pretending that this is the strategy they \enquote{commit} to), and the regions where the other player could remain are the SEC-candidates.
These are then de-/inflated.
We highlight that as the player's estimates change, the SEC-candidates do, too.

\paragraph{Improvements.}
We provide several practical improvements to this approach. 
We briefly describe their ideas here and refer to \ifarxivelse{\Cref{app:3-improvements}}{\cite[App.~C]{pet2-techreport}} for further information.

\begin{compactitem}
	\item Instead of searching SEC-candidates in the complete SG, we once identify all MECs and then search for SEC-candidates in each MEC independently. 
	\item SEC-candidate search is not performed every iteration, but only heuristically (improving a suggestion from~\cite[Sec.~6.2]{EKKW22}).
	\item Instead of computing staying values precisely, which is 
	often unnecessary, we successively approximate them (as suggested in~\cite[App.~E-B]{lics23-techreport}).
	\item If possible, we locally employ MDP solution approaches, \emph{collapsing} ECs that are completely \enquote{controlled} by a single player into one state that then is handled by the normal Bellman updates (as suggested in~\cite[Sec.~6.2]{EKKW22}).
	This transparently generalizes the existing algorithms for MDPs~\cite{atva,DBLP:conf/cav/AshokCDKM17,HM18}.
\end{compactitem}

\section{Partial-Exploration Algorithm for Solving SGs}\label{sec:PE}

\noindent
Here, we present the novel \emph{partial-exploration (PE)} algorithm obtained by combining the \emph{complete-exploration (CE)} algorithm of \cref{sec:CE} with the ideas of partial exploration~\cite{atva,EKKW22,DBLP:conf/atva/Meggendorfer22}.
Intuitively, for particular models and objectives, some states are hardly relevant, and computing their exact value is unnecessary for an $\varepsilon$-precise result.
For example, in the \model{zeroconf} protocol (choosing an IP address in a nearly empty network), it is hardly interesting what we should do when we run into a collision 10 times:
Since this is so unlikely to happen, the exact outcome in this case barely influences the true result.
Thus, we avoid exploring what exactly is possible in this case and just assume the worst.

More generally, we want to avoid working on the complete model when executing Bellman updates or de-/inflating SEC-candidates.
Instead, we use simulations to partially explore the model, finding the states that are likely to be reached under optimal strategies, and focus computations on these.
If the \enquote{relevant} part of the state space (see~\cite{DBLP:journals/lmcs/KretinskyM20}) is small in comparison to the whole model, we can save a large amount of time and memory.
We refer to the~\cite{atva,EKKW22,DBLP:conf/atva/Meggendorfer22} for a comprehensive discussion of the (dis-)advantages of this approach.

\paragraph{Algorithm.}
The algorithm follows the established structure of~\cite{atva,EKKW22}:
We sample a path through the model, at every state picking an action and a successor according to a guidance heuristic that prefers \enquote{relevant} states.
We terminate the simulation when it has reached a state where continuing the path does not generate new information (e.g.\ an EC that cannot be exited).
Then, we perform Bellman updates, but only on the states of the sampled path.
Additionally, we repeatedly identify both collapsible areas and SEC-candidates in the partial model and de-/inflate if necessary.
This final step is the main technical difference to the previous algorithms~\cite{atva,EKKW22}, which only employed collapsing and deflating, respectively.
It also is one of the major engineering difficulties, see \ifarxivelse{\cref{app:PE-engineer}}{\cite[App.~D.1]{pet2-techreport}}.

\paragraph{Soundness and Correctness.}
In \ifarxivelse{\cref{app:PE}}{\cite[App.~D.2]{pet2-techreport}}, we extend the proof of correctness and termination from reachability~\cite[Thm.~3]{EKKW22} to mean payoff.
In the process of proving correctness, we found and fixed an error in the guidance heuristic of~\cite{EKKW22}.

\paragraph{Practical Improvements.}
As before, we applied several optimizations and heuristics to this algorithm.
Broadly speaking, the overall ideas are the same as for the complete exploration approach, however with several intricacies.
In particular, observe that the partial model constantly changes as new states are explored.
Thus, efficiently tracking and updating SEC-candidates or collapsible parts of the game is much more involved and intertwined with the rest of the algorithm.

\section{Tool Description}\label{sec:tool}

\noindent In this section, we briefly describe relevant aspects of \pettwo, focussing on new features and changes compared to its predecessor \petone.
Like \petone, \pettwo is implemented in Java, reads models and objectives specified in the PRISM modelling language, and outputs the computed value in JSON format.

\paragraph{Design Choices.}
Firstly, \pettwo exclusively employs (sound) VI-based approaches, as opposed to, e.g., SI or LP.
It offers two variants, based on complete exploration (CE) and partial exploration (PE), as presented above.

Secondly, \pettwo deliberately comes without any configuration flags.
In contrast, model checkers such as \prism~\cite{prism-games} and \storm~\cite{storm} implement numerous different approaches and variants, each of which offers several hyper-parameters.
While our choice eliminates the potential for fine-tuning, we experienced that even expert users often do not know how to best choose such parameters. 
Thus, we tried to select internal parameters such that the tool works reasonably well out-of-the-box on all models.
This comes with the additional benefit of greatly reducing the number of code paths, which in turn makes testing much easier.



Thirdly, \pettwo fully commits to solving a single combination of model and objective.
This allows us to exploit information about the objective already when building the model; for example, in reachability/safety objectives, we can directly re-map goal and unsafe states to dedicated absorbing states.
Arguably, this might become a drawback when solving multiple queries on the same model, since the model would need to be explored several times.
However, by caching the parts of the model constructed so far, we can effectively eliminate this problem.
We discuss further ways to exploit this design choice in \ifarxivelse{\Cref{app:tool}}{\cite[App.~E]{pet2-techreport}}.

Finally, \pettwo does not differentiate between Markov chains, MDPs, and SGs.
The algorithms are written for SGs, and, whenever possible, apply specialized solutions locally.
For example, if a part of an SG \enquote{looks like} an MDP (because only one player has meaningful choices), PET locally applies MDP reasoning where applicable.
Aside from transparently gaining the performance of these specialized solutions in most cases, this also eliminates code duplication, resulting in a code base that is more understandable, maintainable, and extendable.

\paragraph{Differences to \petone.}
\pettwo effectively constitutes a complete re-write of nearly all aspects of \pet, with roughly 9k lines of code added and 5k deleted compared to \petone (according to \texttt{git}); for comparison, the overall source code of \pettwo has roughly 14k lines.
Most importantly, \pettwo now also parses SGs, fully focusses on solving one model-objective combination, and also provides efficient CE variants of the algorithms.
These CE variants also come with numerous optimizations, such as graph analysis to identify states with value 0 and 1 for reachability and safety, collapsing end components, etc.
Moreover, each PE variant is now specialized to the concrete objective.
\petone tried to unify as many aspects of the sampling approach as possible, which however proved to be a major design obstacle and performance penalty when incorporating all the different specialized solutions and practical optimizations for stochastic games.
Additionally, we also found and fixed a bug in the mean payoff computation of \petone.
In terms of data structures, we improved several smaller aspects of working with probabilistic models.
For example,  the \enquote{standard} internal model representation of \pettwo offers dedicated support for merging / collapsing sets of states while simultaneously tracking predecessors of each state.
We note that the model representation etc.\ is still provided by our separately available, generic purpose library, making many of these improvements available independently of \pet.

\paragraph{Engineering Improvements.}
We evaluated several practical improvements, which sometimes led to quite surprising effects.
We highlight some insights which we deem relevant for other developers.

1)
\enquote{Unrolling} loops in hot zones of the code can lead to significant performance improvements.
For example, to find the optimal action for a Bellman update, we need to use a for-loop to iterate over all available actions.
This process is performed millions of times during a normal execution.
Unrolling and specializing these loops for small action sets ($\leq 3$ actions) led to noticeable performance improvements.
Similarly, switching from for-each loops (which allocate an iterator) to index-based for loops also yielded notable improvements.
	
2)
We trade memory for time by adding additional data structures to optimize certain access patterns.
For example, maintaining the set of predecessors for each state speeds up graph algorithms such as attractor computations. 
Moreover, in several cases it proved beneficial to store information in multiple formats.
For example, on top of a sorted array-based set, we explicitly store the set of successors of a distribution object as a (roaring) bitmap~\cite{DBLP:journals/spe/LemireKK16}, offering fast \enquote{bulk} operations, such as intersection or subset checks. 

%
	
3)
We also investigated ahead-of-time compilation through GraalVM.
While this improved start-up time, it did not result in significant speed-ups but rather even increased the runtime on some of the larger models (even with profile-guided optimization).
We conjecture that this is mainly due to Java's just-in-time compiler being able to apply better fine-tuning.



%
%
\newcommand{\cmpscore}[4]{\ensuremath{#1[#2{+}{#3}/{#4}]}}
\section{Experimental Evaluation}\label{sec:exp}
\noindent We now discuss our evaluation.
The first goal of our experiments is to validate that \pettwo can indeed solve SGs with reachability and mean payoff objectives in a sound way.
Secondly, we assess the impact of our performance improvements and design choices, in particular having only the general algorithm for SG, by comparing \petone and \pettwo.
Finally, we investigate whether our implementation is competitive with other tools.
We report some further insights in \ifarxivelse{\Cref{app:exp-add}}{\cite[App.~F.3]{pet2-techreport}}.
Our artefact, including all data, tools, scripts, logs, etc., is available at~\cite{artifact}.

\subsection{Experimental Setup}\label{sec:exp-setup}

\paragraph{Technical Setup.}
We ran each experiment in a separate Docker container and, as usual, restricted it to a single CPU core (of an AMD Ryzen 5 3600) and 8 GB RAM.
The timeout is 60 seconds (including the startup time of the Docker container).
We ran every instance three times to even out potential fluctuations in execution times.
While the PE approach is randomized by design, even \enquote{deterministic} algorithms may behave differently due to, e.g., non-deterministic iteration order of hash sets.
We observed that the variance is negligible (the geometric standard deviation usually was $\leq 1.05$).
We thus only report the geometric average of the three runs in seconds.
We require an absolute precision of $\varepsilon = 10^{-6}$ for all experiments. 

\paragraph{Metrics.}
To summarize relative performance of \pettwo compared to tool \texttt{X}, we introduce a four-figure score, written \cmpscore{t}{m}{k}{l}, computed as follows:
Let $M$ the set of instances \emph{where both tools terminated in time}.
Then, $t$ equals the geometric mean of $\text{time}_\texttt{X}(I) / \text{time}_{\pettwo}(I)$ over all instances $I \in M$, with $\text{time}_\mathcal{T}$ referring to the overall runtime of tool $\mathcal{T}$ on an instance, $m = |M|$ refers to the number of such instances, while $k$ describes how often \texttt{X} timed out where \pettwo did not and $l$ vice versa.
Note that $t > 1$ indicates that \pettwo is faster on average (on $M$).
When $t < 1$ but $k \gg l$, we see that on instances that both tools solved, \pettwo was slower, but overall, \pettwo solved much more models, which one may still consider advantageous.

\paragraph{Tools.}
Aside from both versions of \pet, for Markov chains and MDPs we consider \prismgames\footnote{Personal communication with the lead developer confirmed that on Markov chains and MDPs \prismgames uses the same approach as \prism.}~\cite{prism-games}, and \storm~\cite{storm}.
On SGs, we compare to the unsound algorithms in \prismgames and \tempest~\cite{tempest} (an extension of \storm), as well as \prismext, an extension of \prismgames with sound algorithms described in~\cite{gandalf,AEKSW22}.
Note that this selection includes all tools that participated in the SG performance comparison in QComp 2023~\cite{QComp23}.
For all tools, we provide the exact version, ways to obtain them, and invocations in \ifarxivelse{\Cref{app:exp-tools}}{\cite[App.~F.1]{pet2-techreport}} and the artefact.

\paragraph{Performance Considerations.}
Restricting to a single CPU is commonly done to ensure that no tool accidentally exploits parallelism.
However, we observed a significant decrease in performance for \pet, even though all algorithms are sequential.
This turned out to be due to garbage collection.
Using \texttt{jhsdb}, we verified that in the single CPU case Java by default selects the Serial GC (instead of the overall default G1GC).
On some instances, we consistently observed improvements of \textbf{up to 33\%} (!), nearly on par with the performance without any CPU restriction, by simply changing to the parallel GC (\texttt{-XX:+UseParallelGC}), even though the parallel GC uses only one thread.
Concretely, comparing CE and PE with Serial and (single-thread) parallel GC, we get scores of 1.06 and 1.04, respectively, meaning that even \emph{on average} this change leads to a significant difference.
Interestingly, for \prismgames the Serial GC performed better.
We configured \pettwo to always use the Parallel GC by default.
In a similar manner, the hybrid engine of \storm experienced a slowdown of more than 30x due to being restricted to a single CPU, which we addressed by adding appropriate switches (see \ifarxivelse{\Cref{app:exp-tools}}{\cite[App.~F.1]{pet2-techreport}} for details).
We are working with the authors of \storm to automatically detect this case.

While these differences would not invalidate our conclusions in particular, we still want to highlight these observations and emphasize the importance of both careful evaluation and choosing good default parameters.


\paragraph{Benchmarks.}
We consider benchmarks from multiple sources.
Firstly, we include applicable models from the quantitative verification benchmark set (QVBS)~\cite{qvbs}, which however does not provide SGs.
Secondly, we consider the SGs used in QComp 2023~\cite{QComp23}.
Finally, we also gather several models from literature, provide variations of existing models, and create completely new models.
For details on the models, 
we refer to \ifarxivelse{\Cref{app:exp-benchmarks}}{\cite[App.~F.2]{pet2-techreport}}.
All models are included in the artefact.

To ease the evaluation, we remove instances of QVBS that are very simple (CE- and PE-approach of \storm and \pettwo taking less than one second) or very time-consuming (all four approaches taking more than 30 seconds).
With such a timespan, differences and trends are clearly visible, but models remain small enough for the experiments to be reproducible within reasonable time.
This filtering reduces the number of executions from nearly 10000 to about 1800.
Even then, 
the overall evaluation still takes about 24 hours (with timeout of 60s).

%

\subsection{Results}

\newcommand{\TO}{80}
\newcommand{\legendmax}{90}
\newcommand{\plotmarksize}{1.5pt}
\newcommand{\axislines}{
		\addplot[black,forget plot,update limits=false] coordinates {(0.0000001,0.0000001) (1000,1000)};
		\addplot[black,dashed,forget plot,update limits=false] coordinates {(0.0000001,0.0000002) (400,800)};
		\addplot[black,dashed,forget plot,update limits=false] coordinates {(0.0000002,0.0000001) (800,400)};
		
		\addplot[gray,dashed,forget plot,update limits=false] coordinates {(0.0000001,\TO) (1000,\TO)};
		\addplot[gray,dashed,forget plot,update limits=false] coordinates {(\TO,0.0000001) (\TO,1000)};
		\addplot[gray,thin,forget plot,update limits=false] coordinates {(60,0.0000001) (60,60)};
		\addplot[gray,thin,forget plot,update limits=false] coordinates {(0.0000001,60) (60,60)};
}
\newcommand{\markincaption}[2]{{\begin{tikzpicture}[baseline=-3pt] \protect\draw[#1,ultra thick,mark size=3pt] plot[mark=#2] (0,0);\end{tikzpicture}}}

\newcommand{\pettable}{\addplot+[only marks,draw=Dark2-A,mark size=\plotmarksize,mark=x,thick]}
\newcommand{\stormtable}{\addplot+[only marks,draw=Dark2-B,mark size=\plotmarksize,mark=star,thick]}
\newcommand{\prismtable}{\addplot+[only marks,draw=Dark2-C,mark size=\plotmarksize,mark=x,thick]}
\newcommand{\prismexttable}{\addplot+[only marks,draw=Dark2-D,mark size=\plotmarksize,mark=asterisk,thick]}

\begin{figure}[t]
	
\begin{minipage}{0.32\textwidth}
\begin{tikzpicture}
\begin{axis}[
		width=\textwidth,height=\textwidth,
		table/col sep=comma,
		x label style={anchor=north,inner sep=0pt},
		y label style={anchor=south,inner sep=0pt},
		xtick={2,10,60}, xticklabels={2,10,60},
		ytick={2,10,60}, yticklabels={2,10,60},
		xmin=1,ymin=1,xmax=\legendmax,ymax=\legendmax, xmode=log,ymode=log,
		axis x line*=bottom,
		axis y line*=left,
		xlabel=\texttt{\petone PE},ylabel=\texttt{\pettwo PE},
	]
	\axislines

	\pettable table [x=pet_1_s, y=pet_2_s] {exp-results/current/csv_pair_1_vs_2_pe.csv};
\end{axis}
\end{tikzpicture}
\end{minipage}
\begin{minipage}{0.32\textwidth}
\begin{tikzpicture}
\begin{axis}[
		width=\textwidth,height=\textwidth,
		table/col sep=comma,
		x label style={anchor=north,inner sep=0pt},
		y label style={anchor=south,inner sep=0pt},
		xtick={2,10,60}, xticklabels={2,10,60},
		ytick={2,10,60}, yticklabels={2,10,60},
		xmin=1,ymin=1,xmax=\legendmax,ymax=\legendmax, xmode=log,ymode=log,
		axis x line*=bottom,
		axis y line*=left,
		xlabel=\texttt{\storm/\prismgames},ylabel=\texttt{\pettwo CE},
	]
	\axislines
	
	\stormtable table [x=Storm, y=pet_2_g] {exp-results/current/csv_pair_ce_vs_storm.csv};
	\prismtable table [x=PRISMe, y=pet_2_g] {exp-results/current/csv_pair_ce_vs_prism_e_mdp.csv};
\end{axis}
\end{tikzpicture}
\end{minipage}
\begin{minipage}{0.32\textwidth}
\begin{tikzpicture}
\begin{axis}[
		width=\textwidth,height=\textwidth,
		table/col sep=comma,
		x label style={anchor=north,inner sep=0pt},
		y label style={anchor=south,inner sep=0pt},
		xtick={2,10,60}, xticklabels={2,10,60},
		ytick={2,10,60}, yticklabels={2,10,60},
		xmin=1,ymin=1,xmax=\legendmax,ymax=\legendmax, xmode=log,ymode=log,
		axis x line*=bottom,
		axis y line*=left,
		xlabel=\texttt{\tempest/\prism variants},ylabel=\texttt{\pettwo CE},
	]
	\axislines
	
	\prismtable table [x=PRISMe, y=pet_2_g] {exp-results/current/csv_pair_ce_vs_prism_e_smg.csv};
	\prismexttable table [x=PRISM-extensions, y=pet_2_g] {exp-results/current/csv_pair_ce_vs_prism_ext.csv};
	\stormtable table [x=TEMPEST, y=pet_2_g] {exp-results/current/csv_pair_ce_vs_tempest.csv};
\end{axis}
\end{tikzpicture}
\end{minipage}

\caption{
	Comparison of \pettwo to other tools.
	From left to right we compare {\pettwo}-PE and {\petone}-PE, {\pettwo}-CE on MDP and MC with \storm (\markincaption{Dark2-B}{star}) and \prismgames (\markincaption{Dark2-C}{x}), and finally \pettwo-CE on SG with \tempest (\markincaption{Dark2-B}{star}), \prismgames (\markincaption{Dark2-C}{x}), and \prismext (\markincaption{Dark2-D}{asterisk}).
	A point $(x, y)$ denotes that tool X and \pettwo needed $x$ and $y$ seconds, respectively.
	If a point is above/below the diagonal, tool X is faster/slower.
	Plots are on logarithmic scale, dashed diagonals indicate that one tool is twice as fast.
	Timeouts are pushed to the orthogonal dashed line.
}
\label{fig:results}
\end{figure}
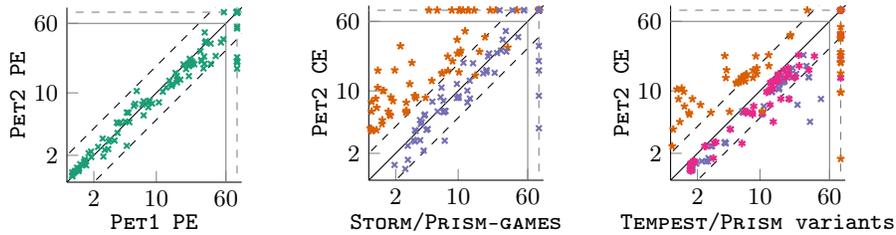

\noindent
We present central results in \Cref{fig:results} and discuss each of our research questions.

\paragraph{Soundness and Scalability.}
We empirically validate the correctness of \pettwo by (i)~comparing against the reference results in QVBS (this only affects the specialized MDP reasoning of our algorithm), (ii)~ensuring that both algorithms inside \pettwo yield the same results, and (iii)~comparing against manually computed values, both for existing SG benchmarks as well as handcrafted ones exhibiting various graph structures (some of which arose as test or corner cases).
In all cases, \pettwo's results are sound, i.e.\ within the allowed precision of $\varepsilon=10^{-6}$.
In contrast, throughout the whole SG benchmark set, \prismgames and \tempest return several wrong answers, see \ifarxivelse{\Cref{app:exp-add}}{\cite[App.~F.3]{pet2-techreport}} for details.
In particular, \tempest returns wrong answers in 6 out of 13 cases where we have known reference results, often by a significant margin, e.g.\ returning 0.0003 instead of 0.481.

Additionally, we see that \pettwo can solve models with millions of states and various difficult graph structures within a minute.
Thus, we conclude that both our algorithm and implementation scale well.

\paragraph{Comparison to \petone.}
When solving Markov chains or MDPs, \pettwo still uses algorithms that can handle SGs.
This generality comes with some overhead, for example because data structures for tracking ownership of states are not necessary in MDPs.
However, a score of \cmpscore{1.07}{85}{8}{1} and \Cref{fig:results} (left) show that PE in both versions of \pet performs remarkably similar, with \pettwo even slightly faster. 
We conclude that the improvements in the implementation make up for the algorithmic overhead.



\paragraph{Comparison to other Tools.}
It is well known that the \emph{structure} of a model is \emph{the} determining factor for the relative performance of different algorithmic approaches, see e.g.~\cite{AEKSW22,DBLP:conf/atva/Meggendorfer22,DBLP:conf/tacas/HartmannsJQW23}, in particular far more than the number of states or transitions.
(This is also supported by our comparison of CE and PE in \ifarxivelse{\Cref{app:exp-add}}{\cite[App.~F.3]{pet2-techreport}}, sometimes showing order of magnitude advantages in either direction.)
Thus, instead of comparing tools as a whole, we compare matching algorithmic approaches (CE and PE based value/interval iteration) to assess only the impact of different implementations.
For similar reasons, here we only compare the explicit engines and not symbolic or hybrid approaches (results on the latter are provided in \ifarxivelse{\Cref{app:exp-add}}{\cite[App.~F.3]{pet2-techreport}}).

Comparing PE approaches, \pettwo outperforms the only competitor \storm with a score of \cmpscore{0.3}{27}{29}{1} (\pettwo solves more than twice the number of models); see \ifarxivelse{\Cref{app:exp-add}}{\cite[App.~F.3]{pet2-techreport}} for details.
For CE algorithms, across all instances where both tools are applicable, the score of \pettwo against the Java-based tools \prismgames and \prismext is \cmpscore{1.3}{104}{10}{4} and \cmpscore{1.3}{53}{4}{0}, respectively, while against the C++ tools \storm and \tempest it achieves \cmpscore{0.3}{69}{0}{12} and \cmpscore{0.4}{54}{13}{1}, respectively.
For a more detailed comparison, \Cref{fig:results} (middle) compares the CE algorithms of \pettwo with those in \storm and \prismgames on Markov chains and MDPs.
Here, as expected, \storm outperforms other tools, 
at least partially due to performance differences of C++ and Java.
However, \pettwo performs favourably against the state-of-the-art tool \prismgames. 
This shows that our first, generic implementation of the CE algorithm for SG is comparable to established tools even on Markov chains and MDPs. 
Finally, \Cref{fig:results} (right) compares \pettwo with the other tools on SGs, namely \prismgames, \tempest, and \prismext.
Recall that only \prismext uses a sound algorithm, while the other tools use an unsound stopping criterion and thus require less work. 
Nonetheless, \pettwo often outperforms the other tools (even \tempest, which builds on the highly optimized \storm), making it the most viable tool on SGs not only because of soundness, but also because of performance.

Finally, to (superficially) evaluate how much objective-specific optimizations yield, we implemented viewing reachability objectives as \enquote{trivial} mean payoff objective, i.e.\ goal states are set to be absorbing with reward 1, and all others with reward zero.
This modified query is then passed to our generic mean payoff algorithm.
Notably, even then \pettwo slightly outperforms \prismgames solving the reachability objective directly (\cmpscore{1.1}{98}{8}{10}), and in turn the dedicated reachability approach of \pettwo \enquote{only} scores \cmpscore{1.2}{106}{8}{0} against this variant.

\section{Conclusion}

\noindent We presented \pettwo, the first tool implementing a sound and efficient approach for solving SGs with objectives of the type reachability/safety and mean payoff.
Our experimental evaluation shows that (i)~it is sound, while other tools indeed return wrong answers in practice, (ii)~it offers the most efficient partial-exploration based algorithm, and (iii)~it is the most viable tool on SGs.
%

For future work, there is still a lot of room for heuristics and engineering improvements, for example adaptively choosing internal parameters, more efficient tracking and handling of SEC-candidates, using topological order of updates in VI, improved pre-computation for mean payoff, etc.
Additionally, support for total reward is planned; however, as described in \ifarxivelse{\Cref{app:TR}}{\cite[App.~B]{pet2-techreport}}, this requires using ideas such as optimistic value iteration~\cite{DBLP:conf/cav/HartmannsK20,AEKSW22} in order to be reasonably efficient.

\newpage

%

%
%
%
\bibliographystyle{splncs04}
\bibliography{ref}

\ifarxivelse{
\newpage
\appendix
\section{Extended Preliminaries}\label{app:prelims}

\noindent Here, we provide extended version of the preliminaries, complementing (and, for the sake of readability, sometimes duplicating) the information in \Cref{sec:prelims}.
This section is based on the preliminaries in the theoretical basis of our algorithms~\cite{lics23}.

$\Distributions(X)$ denotes the set of all \emph{probability distributions} over a countable set $X$, i.e.\ mappings $\distribution : X \to [0, 1]$ such that $\sum_{x \in X} \distribution(x) = 1$.
For a set $S$ we denote by $S^*$ and $S^\omega$ the set of finite and infinite sequences of elements of $S$, respectively.

A \emph{(turn-based) stochastic game (SG)} (e.g.\ \cite{condonAlgo}) is a tuple $(\states,\maxstates, \minstates, \act, \trans)$, where $\allstates$ is a finite set of states partitioned into $\maxstates$ and $\minstates$, the sets of states belonging to the \emph{Maximizer} player $\maxsymbol$ and the \emph{Minimizer} player \raisebox{0.07cm}{$\minsymbol$}, respectively;
further, $\act$ denotes a finite set of \emph{actions} and we overload $\act$ to also act as a function assigning to each state $s$ a non-empty set of \emph{available actions} $\act(s)$; and
$\trans : \allstates \times \act \to \Distributions(\allstates)$ is the \emph{transition function} that for each state $s$ and (available) action $a \in \act(s)$ yields a distribution over successor states.
An SG where all states belong to one player is called Markov decision process (MDP), see~\cite{DBLP:books/wi/Puterman94}.


The \emph{semantics} are defined as usual: Choices are resolved by strategies, inducing a Markov chain with the respective probability space over infinite paths; we shortly recall this approach in the following and refer to~\cite[Chap. 2]{MaxiThesis} for a more extensive description.
Intuitively, a stochastic game is played in turns:
In every state $s$, the player to whom it belongs chooses an action $a$ from the set of available actions $\act(s)$ and the play advances to a successor state $s'$ according to the probability distribution given by $\trans(s, a)$.
Starting in a state $s_0$ and repeating this process indefinitely yields an infinite path $\infinitepath = s_0 a_0 s_1 a_1 \dots \in (\allstates \times \act)^\omega$ such that for every $i \in \Naturals_0$ we have $a_i \in \act(s_i)$ and $s_{i+1} \in \{s' \in \allstates \mid \trans(s_i, a_i)(s') > 0\}$.
$\Infinitepaths<\G>$ denotes the set of all such infinite paths in a given game $\G$.
Furthermore, we write $\infinitepath_i$ to denote the $i$-th state $s_i$ in a path.

A \emph{memoryless deterministic (MD) strategy} assigns a chosen action to every state. Formally, an MD strategy of Maximizer $\straa : \maxstates \to \act$ or Minimizer $\strab : \minstates \to \act$ is a function mapping all states of the player to an available action, i.e.\ $\straa(s) \in \act(s)$ for all $s$.
Resolving all nondeterministic choices by fixing a pair of strategies $(\straa,\strab)$, we obtain a Markov chain $\G^{\straa, \strab}$.
This Markov chain together with a state $s$ induces a unique probability distribution $\probability_{\G, s}^{\straa, \strab}$ over the set of all infinite paths $\Infinitepaths<\G>$ \cite[Sec.~10.1]{DBLP:books/daglib/0020348} (where the set of paths starting in $s$ has measure 1).
For a random variable over paths $X : \Infinitepaths<\G> \to \Reals$, we write $\expec_{\G,s}^{\straa,\strab}[X]$ for the expected value of $X$ under the probability measure $\probability_{\G, s}^{\straa, \strab}$.

An \emph{objective}  $\Phi : \Infinitepaths<\G> \to \Reals$ formalizes the \enquote{goal} of both players by assigning a value to each path.
In this paper, we focus on \emph{mean payoff} (also called long-run average reward)~\cite{gillette1957stochastic} objectives.
These are based on a reward function $\statereward : \allstates \to \Reals$ which assigns real numbers to all states.
The mean payoff of a path is the average reward obtained in the limit, formally $\Phi(\infinitepath)
\eqdef \liminf_{n \to \infty} (\frac{1}{n} \sum_{j=0}^{n-1} \statereward(\rho_j))$.

The presented algorithms and tools can also explicitly handle reachability and safety objectives.
Intuitively, these compute the probability of reaching a set of goal states, or of avoiding a set of unsafe states.
We omit the formal definition, since abstractly both objectives are degenerate cases of mean payoff, see e.g.~\cite{DBLP:conf/cav/AshokCDKM17}.
Additionally, another prominent objective is known as \emph{total reward}, which we separately discuss in \Cref{app:TR}.

The two players are antagonistic, and, as their names suggest, Maximizer aims to maximize the obtained value, while Minimizer wants to minimize it (i.e.\ the game is zero-sum).
We are interested in the \emph{value of the game}, i.e.\ the optimal value the players can ensure.
Formally, the value of state $s$ is defined as
$
\val_{\G, \Phi}(s) \eqdef {\sup}_\straa~{\inf}_\strab~\expec_{\G, s}^{\straa, \strab}[\Phi] = {\inf}_\strab~{\sup}_\straa~\expec_{\G, s}^{\straa, \strab}[\Phi],
$
where the latter equality holds for all objectives we consider \cite{DBLP:journals/jsyml/Martin98,DBLP:journals/ijgt/MaitraS98,DBLP:conf/lics/KieferMSW17a}, i.e.\ the games are \emph{determined}.
Moreover, MD strategies are sufficient for playing optimally~\cite[Thm.~1]{liggett1969stochastic}.
We are interested in \emph{approximate solutions}, i.e.\ given a concrete state $\initstate$ and precision requirement $\varepsilon$, our goal is to determine a number $v$ such that $|\val_{\G, \Phi}(\initstate) - v| < \varepsilon$.

\emph{End components} intuitively are sets of states in which the system can remain forever, given a suitable pair of strategies.
Formally, a pair $(R, B)$, where $\emptyset \neq R \subseteq \allstates$ and $\emptyset \neq B \subseteq \Union_{s \in R} \act(s)$, is an \emph{end component (EC)}~\cite{dA97} if
(i)~for all $s \in R$ and $a \in \act(s) \intersection B$ we have $ \{s' \in \allstates \mid \trans(s, a)(s') > 0\}\subseteq R$, and
(ii)~for all $s, s' \in R$ there is a finite path $s a_0 \dots a_n s' \in (R \times B)^\star \times R$, i.e.\ the path stays inside $R$ and only uses actions in $B$.
Inclusion-maximal ECs are called \emph{maximal end component (MEC)}.
MECs are an essential object for SG, since the play of an SG eventually remains inside a single MEC with probability one~\cite{dA97}.
In other words, MECs capture the possible long run behaviour of the system.

\section{A Note on Expected Total Reward Objectives}\label{app:TR}

\noindent We define total reward objectives, because they are included in the theoretical basis of this work~\cite{lics23} and we comment why we chose not to include them in our tool yet.
Like mean payoff objectives, total reward objectives use a reward function, but instead of the average, they consider the sum of rewards collected along a path, i.e.\  $\Phi(\infinitepath) = {\sum}_{i=0}^\infty \statereward(\infinitepath_i)$.
This sum can be infinite, and need not be defined if the reward function assigns both positive and negative rewards~\cite[Sec.~5.2]{DBLP:books/wi/Puterman94}.

First, for partial exploration, observe that we cannot derive an a-priori upper bound on the value as we could with, e.g., reachability.
In particular, we even cannot be sure whether there are states which provide infinite reward.
Thus, no matter how unlikely it is to reach a particular state, we have to determine its value -- if it is infinite, the overall value immediately is infinite.
Thus, without a guarantee that the reward is not infinite, partial exploration cannot yield any of its advantages because the whole model has to be explored.

Secondly, even if we consider complete exploration and exclude all states with infinite reward, the only easily derivable a-priori upper bound on the reward is $\mathcal{O}(p_{\min}^{-|\states|} \cdot r_{\max})$, where $p_{\min}$ is the smallest non-zero transition probability and $r_{\max}$ the largest occurring reward, cf.~\cite[Sec. 3.3]{CFK+13reward}.
The leading factor $p_{\min}^{-|\states|}$ is a bound on the so-called \emph{mixing rate} of the system.
In a nutshell, this effectively describes the \enquote{speed} with which we could move through the system.
(Technically, this rate is related to the second eigenvalue of the transition matrices of the induced Markov chains which in turn gives the convergences rate of the power method, i.e.\ value iteration.)
Even for MDP and MC, precisely determining this factor is known to be as hard as solving the problem itself (PTIME).
We also are not aware of any approach of deriving a significantly better a-priori bound in practice, i.e.\ one that quickly yields non-exponential bounds on many models.
(In the following, we refer to works that bound the mixing rate based on the topological structure of the system, however the worst case still is exponential.)

To show the impact of this upper bound, consider the following construction:
Let $\MC$ be any Markov chain with a single-state reachability objective, say state $g$.
Let $n$ be the number of steps that VI needs to converge until precision $\varepsilon$.
Omitting some technical details, for sufficiently complex Markov chains, $n$ strongly correlates with the concrete mixing rate $\gamma$, which intuitively indicates the convergence rate of the iteration, see e.g.\ \cite[Sec. 3.5]{DBLP:conf/spin/ChatterjeeH08} or~\cite[Sec. 3.3]{HM18}. 
In particular, we have that $n \approx \log \varepsilon / \log \gamma$ (where $\gamma$ close to 1 indicates that the MC is \enquote{slowly mixing}, sometimes called \emph{stiff}, and VI takes long to converge).

Now, suppose that we modify the Markov chain as follows:
We let $g$ transition to a fresh sink state and give all states a reward of $0$, except $g$ which receives a value of $1$.
Naturally, the maximal total reward in this system equals the reachability probability.
However, if we were to apply total reward iteration on this system together with the naive upper bound from above, we would not start with an upper bound of $1$ (as with reachability), but $\mathcal{O}(p_{\min}^{-|\states|})$.
Glossing over details, recall that $\gamma$ corresponds to the convergence rate, i.e.\ after $k$ steps we expect the difference between lower and upper bound to have decreased by $\gamma^k$.
So, to get to a precision of $\varepsilon$, we now need $\approx (|\states| \cdot \log(p_{\min}) \cdot \log \varepsilon) / \log \gamma$, i.e.\ $|\states| \cdot \log(p_{\min}) \cdot n$.

Naturally, simple graph analysis could derive that $1$ is a tighter upper bound on this particular example.
However, as mentioned above, we are not aware of any methods that provide good a-priori bounds in the general case.
Then, the conclusion that total reward iteration may require on the order $|\states| \cdot \log(p_{\min})$ more steps than reachability still holds.
Using a preliminary implementation, we also confirmed the problem in practice.

So how could one avoid these problems and make value iteration performant when facing total reward objectives?
One can improve the methods for getting an upper bound; in particular, we can exploit the topology of the model (e.g.\ analysing strongly connected components) to decrease the exponent of $p_{\min}$. While this might help in practice, in theory it cannot solve the problem, as there are models where there is no better upper bound, similar to the one in~\cite[Fig. 3]{HM18}.

Orthogonally, in~\cite{DBLP:conf/cav/HartmannsK20}, the authors introduce a variant of value iteration that first only performs Bellman updates on an under-approximation until the estimates do not change by much (the \enquote{naive} stopping criterion).
Then, it \enquote{guesses} a value vector that is slightly larger and checks whether this is a proper upper bound. 
If it is, the algorithm can immediately terminate; otherwise, the upper bound candidate is discarded and the value iteration from below continues with a higher precision requirement.
This effectively eliminates the need to have an a priori upper bound.
This so-called \emph{optimistic value iteration} has also been extended to SGs with reachability objectives~\cite{AEKSW22}, where the verification of the upper bound is technically more involved.
We plan to implement optimistic value iteration in \pet.
The results of~\cite[Sec. 5.2]{DBLP:conf/cav/HartmannsK20} suggest that optimistic value iteration is a lot more effective on MDPs with total reward in practice.

%
%
\section{Details on Practical Improvements}\label{app:3-improvements}

\paragraph{Simplified Mechanism for Detecting SECs.}
%
%
%
%
%
%
%
%
As briefly mentioned in the main body, we first search for MECs in the overall SG.
Since SECs can only be subsets of a MEC, whenever we search for SEC-candidates, we then can restrict to each MEC of the SG individually.
Similarly, we only need to track the the \emph{recommender strategy} (a technical device of \cite{lics23} used to identify SECs) within each MEC of the game.

While in the CE algorithm, the set of MECs of the SG can be determined beforehand, this is complicated in the PE case, where we repeatedly have to search for components in the (partially explored) SG:
By adding states, the ECs we identify in the partial system can grow or even merge.
Thus we repeatedly need to update the game components, within those search for SEC-candidates, and, ideally, carry over any information on the recommender strategy to the grown components instead of starting from scratch.

\paragraph{Deciding when to Search for SECs.}
While the algorithm of \cite{lics23} searches for SEC-candidates and applies de-/inflation in every step, both correctness and termination proof only require that these steps happen infinitely often.
Thus, instead of applying this whole process in every step of the algorithm, we lazily delay these updates.

In~\cite[Sec.~6.2]{EKKW22}, the authors suggest to perform the search and de-/inflation every $n$ steps.
We improve on this suggestion by making the SEC-search depend on the evolution of the strategies during the execution of the algorithm.
Recall that we track SEC-candidates individually per MEC.
Since SEC-candidates remain the same until the recommender strategy changes, we only mark a SG-MEC \enquote{stale} once an optimal action changes.
Then (potentially after a few further steps) a SEC-search is performed in this MEC.
This means that as long as the optimal strategies inside MECs stay the same, we completely avoid having to search SECs.

%
%

\paragraph{Successively Approximating Staying Values.}
We implemented most ideas listed in \cite[App.\ E-B]{lics23-techreport}.
Concretely, instead of computing the precise staying value by, e.g., solving an LP every time, we run a value iteration, approximating the mean payoff and use it to derive (converging) bounds on the staying value.
Similarly, whenever we identify new SEC-candidates, we re-use the results of the previous iteration.
This is particularly important for the PE approach, where the set of SEC-candidates frequently changes.

\paragraph{Collapsing State Sets where Possible.}
%
%
Collapsing a set of states means that we replace it with a single representative; this is possible when we can prove that all states in the set have the same value.
This replacement is useful because (i) we simplify the model, and (ii) instead of repeatedly de-/inflating the whole set, we can just apply Bellman updates on the representative.

To collapse state sets in SGs with mean payoff objectives, we combine the ideas of~\cite{DBLP:conf/cav/AshokCDKM17} (collapsing for mean payoff objectives in MDPs) and~\cite{EKKW22} (collapsing for reachability objectives in SGs).

We use several approaches to identify collapsible state sets:
Firstly, in CE we can identify all states in the \emph{attractor} of goal and sink regions by graph analysis, which have a value of 1 and 0 respectively.
Moreover, in the process of solving a game, we may derive for additional states that they have, e.g., a value of 1.
We repeatedly perform an attractor computation for all states with value 0 and 1.
Secondly, we search for \emph{controlled ECs}, i.e.\ ECs in the game where only one player has any meaningful choice and thus controls whether to remain inside the EC or choose any of the available exists.
Such ECs form equivalence classes w.r.t.\ the value and thus can be collapsed.
Note that this extends the classical idea of MEC-quotient~\cite{DBLP:conf/cav/AshokCDKM17} to games; moreover, it extends the idea of collapsing ECs belonging fully to one player~\cite{MatPhD}.

In the future, we want to investigate whether we can easily identify further regions to collapse.
For example, if we can show that for a general MEC in an SG, i.e.\ where both players make decisions, that under any optimal strategy the MEC is left, we can again remove that region.
Similarly, any EC where all states provably achieve the same value (i.e.\ the upper and lower bound of all states equal the same value), can be summarized to a single state.
%

\section{Details on the Partial-Exploration Approach}

\subsection{Explanation of the Engineering Difficulties}\label{app:PE-engineer}

We provide an intuitive description of the emerging engineering difficulties.
A full technical explanation would require detailed knowledge of~\cite{lics23} (in particular, how SEC-candidates emerge), with details unfortunately going beyond the scope of this paper.

In essence, PE on MDP could simply repeatedly search for ECs, merge each found component into a representative, and (mostly) forget about all internal structure.
For SGs, we search for ECs in two "layers".
First, we search ECs \emph{in the game}, since any SEC is subset of ECs in the game.
For such game-ECs, we check whether they are \enquote{controlled} by one player and, if yes, apply MDP-based reasoning if possible.
Otherwise, for such an EC, we initialize tracking objective-specific information to derive recommended strategies and heuristics when to search for SECs.
(For example, we apply the total reward iteration required for mean payoff on such a game EC.)
Note that without identifying these game-ECs, we would need to track this information globally!
Then, within these game ECs, we repeatedly need to search for SEC candidates (EC search in induced MDP) and compute/approximate their staying value.
For this process, we need to re-use information computed in the outer \enquote{tracking} EC.
Then, deflating and inflating of values happens on the game layer, however using values computed inside the SEC candidates.
On top of all this, since we are sampling and dynamically building the game, we also may need to grow or merge game SECs \emph{without discarding previously gained information}.

To summarize, the difficulty is the need to operate on multiple different layers (overall game, induced MDPs, staying value within changing SEC-candidates...), with information flowing between these layers in both directions. 



\subsection{Proof of Correctness and Termination of the Partial-Exploration Algorithm}\label{app:PE}

\noindent We now prove that the new partial-exploration algorithm described in \Cref{sec:PE} is correct and terminates for all precision requirements $\varepsilon>0$.

\paragraph{Correctness.}
The algorithm starts from correct estimate vectors $\lb$ and $\ub$ and then applies two operations: Bellman updates and de-/inflating of SEC-candidates.
By a simple induction, Bellman updates always yields a correct under-/over-approximation again, also when only executed on a subset of states.

For the correctness of de-/inflating, we focus only on deflating, as the other case is analogous.
Let $(R,B)$ be a SEC-candidate in the SG.
This means that, in our current partial model of the SG, if we fix the locally optimal Minimizer strategy, $(R,B)$ is a MEC in the resulting partial model of the MDP.
We emphasize the partial model aspect because this is the crucial difference to~\cite{lics23}: 
$(R,B)$ might not be a MEC.

Observe that we only apply deflating on SEC-candidates. 
A SEC-candidate is a MEC in the MDP that is obtained after fixing a Minimizer strategy.
We now state two facts:
1. The value of all states in this MDP is greater or equal than their value in the game, as Minimizer's strategy might have been suboptimal (see~\cite[Eq. (4)]{lics23}).
2. Every EC in an MDP is also a SEC in this MDP~\cite{EKKW22}.
Thus, by~\cite[Lem. 5]{lics23}, executing deflate on this SEC yields a value that is greater or equal than the value of the states in the MDP.
Together, these facts imply that deflating is correct.

We highlight an intricate difference to the proof in~\cite{lics23}: The SEC-candidate is only a MEC in the \emph{partial model} of the MDP, but it might actually be included in some larger EC when considering the full model.
Nonetheless, our proof only requires that it is an \emph{EC} in the MDP, not that it is inclusion maximal.

\paragraph{Termination without Guidance Heuristic.}
To simplify the proof of termination, first assume that we do not employ any guidance heuristic, but rather choose actions at random and sample successor states of a state-action pair according to the distribution.
This way, if the algorithm runs forever, almost surely every state of the SG is seen infinitely often.
Thus, every state is also updated by Bellman updates infinitely often, the algorithm explores and constructs the complete model, and it executes de- and inflating on all SEC-candidates infinitely often.
Thus, by~\cite[Thm. 3]{lics23}, its precision converges to 0, and for all $\varepsilon>0$, it terminates.

\paragraph{Error in the Guidance Heuristic of~\cite{EKKW22}.}
Before we prove that we can employ guidance heuristics, we show how a guidance heuristic might make the algorithm non-terminating.
In~\cite{EKKW22}, the authors apply the same idea as in the partial-exploration algorithm for MDPs~\cite{atva}: pick the action with the highest upper bound for Maximizer states or the lowest lower bound for Minimizer states.
However, in MDPs, ECs are collapsed and thus do not affect the sampling after being detected.
In contrast, in SGs, ECs are only deflated. 
Moreover, in~\cite{EKKW22} SEC-candidates are only searched along the path that was sampled, not in the whole partial model.
To show why this guidance heuristic does not suffice to terminate, consider the MDP in \Cref{fig:guidance-counter-example}.
We restrict to this MDP for simplicity, but assume we do not collapse but only deflate. 
By adding Minimizer states to the cycles, we can construct a similar example as an SG where collapsing cannot be applied.

The objective is to reach state $t$, and we start with $\ub(s_1)=\ub(s_2)=1$.
As initially all actions yield the same upper bound, eventually our simulation reaches either $t$ or $z$. 
Reaching $t$ does not affect the upper bounds, as it is the target state and has an upper bound of $1$.
In contrast, reaching $z$, we determine (by deflating) that it has an upper bound of $0$. 
Thus, when first reach $z$, we also deflate $s_2$ and set its upper bound to $\ub(s_2, b_2)=\frac 2 3$;
and we deflate $s_1$ and set its upper bound to $\ub(s_1, b_1)=\frac 4 6$.

Now, the next simulation will be stuck in $s_1$, since it has a higher upper bound than $s_2$. 
Deflating the SECs along the path (i.e.\ only $\{s_1\}$) does not change that, because no matter how often we deflate, we will always have $\ub(s_1)>\ub(s_2)$.
Thus, with such a chain of SECs, the partial-exploration algorithm as in~\cite{EKKW22} does not terminate.

\begin{figure}[t]
	\centering
	\begin{tikzpicture}[initial text=\empty]
		\node[max vertex,initial=left] at (-0.5,0) (s1) {$s_1$};
		\node[max vertex] at (2,0) (s2) {$s_2$};
		\node[actionnode] at (0.75,0) (mid1) {};
		\node[actionnode] at (3,0) (mid) {};
		\node[max vertex] at (4.5,0.5) (t) {$t$};
		\node[max vertex] at (4.5,-0.5) (z) {$z$};
		
		\path[->,directedge]
		(s1) edge[loop above] node[anchor=south,action] {$a_1$} (s1)
		(s2) edge[loop above] node[anchor=south,action] {$a_2$} (s2)
		(t) edge[loop right] (t)
		(z) edge[loop right] (z)
		;
		\path[-]
		(s2) edge node[anchor=north,action] {$b_2$} (mid)
		(s1) edge node[anchor=north,action] {$b_1$} (mid1)
		;
		\path[probedge]
		(mid) edge node[prob,above] {\small $\frac 1 3$} (t)
		(mid) edge node[prob,below] {\small $\frac 1 3$} (z)
		(mid) edge[out=45,in=45,looseness=1.5] node[prob,above,pos=0.6] {\small $\frac 1 3$} (s2)
		(mid1) edge[out=45,in=45,looseness=1.5] node[prob,above,pos=0.6] {\small $\frac 1 2$} (s1)
		(mid1) edge node[prob,below,sloped] {\small $\frac 1 2$} (s2)
		;
	\end{tikzpicture}
	\caption{Example MDP to show the error in the guidance heuristic of~\cite{EKKW22}.}
	\label{fig:guidance-counter-example}
\end{figure}
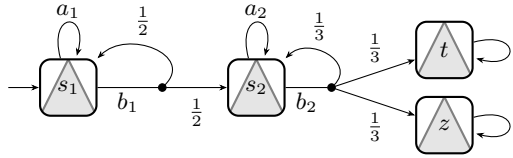

On a high level, the problem is that there is a SEC with a dependence on itself, so that it always seems optimal to stay, even after deflating. 
One way of addressing this is to remove such simple self-loops and redistribute the probability among the other successors (in the example, $b_1$ surely goes to $s_2$ and $b_2$ goes to $t$ or $z$ with probability $\frac 1 2$ each).
This solves this particular example, and has the additional advantage that it speeds up the convergence. 
However, it does not suffice in general, because the EC can be larger and with non-trivial structure; then, detecting and redistributing all transitions leading back to the EC is not possible.
Instead, our algorithm remembers all SEC-candidates that were deflated, and upon reaching any state in this SEC jumps to the exit that was used for the deflation. 
This way, our simulations are not stuck.
(We remark that a similar idea was discussed in~\cite[Sec. 6.2]{EKKW22}, but deemed inefficient because of the prototypical implementation.)

\paragraph{Termination with guidance heuristic.}
Finally, we prove that the algorithm terminates even with the guidance heuristic.
Previously, we used the fact that all states were visited and updated infinitely often.
However, with the guidance heuristic, some states are eventually not visited any more.
This is a key feature of partial exploration: when we can prove that some action is suboptimal, we can avoid following it and exploring the states behind it, see~\cite[Ex. 9]{EKKW22}.

To prove that the algorithm still terminates, intuitively, we instead use the assumption that all actions that are optimal in the limit are played infinitely often, and thus all states reachable by an optimal strategy are reached infinitely often.
For a more formal description formulation of this assumption, we refer to~\cite[Sec.~4.1.1]{DBLP:conf/concur/GroverKMW22}. 
Let $\states_\infty$ by the states that are updated infinitely often.
If $s\notin \state_\infty$, then it is not reachable under optimal strategies; this means that every action that leads to $s$ becomes suboptimal at some point during the execution of the algorithm. Increasing the precision in $s$, i.e.\ lowering its upper bound, cannot change the fact that it is suboptimal; so the fact that we only update $s$ finitely often (often enough to prove its suboptimality) does not affect convergence or termination.
If $s\in\states_\infty$, it is updated infinitely often. Thus, on the SG restricted to $\states_\infty$ we can apply the reasoning above, utilize~\cite[Thm. 3]{lics23} and thereby prove that the values converge in the limit and thus, for every $\varepsilon>0$, the algorithm terminates.

\section{Restricting to Single Model-Property Combinations}\label{app:tool}

\noindent
As mentioned in the main body, \pettwo restricts to solving one concrete model-property combination.
For PE approaches, this is fundamentally required, since information about the property is required for guidance and the interleaving of exploration and solving.
We deliberately carry over this decision to the CE case, as this analogously allows us to exploit information about the objective directly while building the model.
In particular, for reachability/safety objectives, instead of building the whole model state-by-state, we initially create a dedicated goal and sink state and then, while exploring the model, immediately replace any state that matches the reachability condition / violates the safety condition with the dedicated goal and sink state, respectively.
This optimization allows us to potentially omit constructing large parts of the state space.

For a more technical, yet practically important optimization, we briefly explain the input format in which games are provided, namely the \prismgames modelling language, see \url{http://prismmodelchecker.org/games/modelling.php}.
This language also allows to specify, e.g., concurrent stochastic games or non-zero sum properties etc.
As such, it allows to declare multiple players (i.e.\ more than two).
To obtain the kind of zero-sum, two-player games we are interested in, the \emph{properties} specify a \emph{coalition} of players which together should act as the Maximizer player, playing against all other players which together act as Minimizer.
Now, if we would build and store the game without this coalition information, we would need to track for each state / action the associated player and then, once we receive the property, either \enquote{project} out the previously stored information or, even worse, whenever we need to know which player a state belongs to check whether its owner is part of the Maximizer coalition or not. (This is exactly what \prismgames does.)
In contrast, by combining model and property, we can directly track which states are \enquote{owned} by Maximizer and Minimizer in corresponding sets.
(Internally, \pettwo stores which states are \emph{decision} states of either player, i.e.\ states which are (i)~owned by the respective player and (ii)~have more than one action available, which is used when identifying controlled ECs mentioned above.)

A potential disadvantage of this approach is that if we were to solve multiple properties on the same model, \pettwo potentially has to explore it completely multiple times.
This could be partially alleviated by caching which part of the underlying model we explored in an \enquote{unmodified} way and successively building up this cache when needed.
However, \pettwo currently does not implement such a cache, mainly for the sake of simplicity.
In particular, current benchmarking setups evaluate every model-objective combination separately.

%
\section{Additional Information on Evaluation}

\subsection{Tools}\label{app:exp-tools}

\noindent For each tool, we provide the version, ways to obtain it and a related citation, together with the exact invocations used in our evaluation.

\begin{description}
	\item[\pettwo]
	Our own tool, presented in this paper, available at:
	\begin{center}
		\url{https://gitlab.lrz.de/i7/partial-exploration}
	\end{center}

	For the invocations, the precision requirement is set by specifying \texttt{-{}-precision 1e-06} (which also is the default value).
	We omit mentioning this switch below for readability.
	The model is specified as \texttt{-{}-model <model file> -{}-const <constants>}.
	The property is specified by \texttt{-{}-properties <properties file> -{}-property <property name>}.
	Once \pet natively supports parsing the PRISM language (i.e.\ without relying on PRISM), we will add support for specifying mean payoff in the property file.
	\begin{itemize}
		\item Reachability: \texttt{prism <model> <property>}

		PE by default, add \texttt{-{}-global} to switch to CE

		\item Mean payoff: \texttt{mean-payoff <model> -{}-reward <reward name>}
		\begin{itemize}
			\item PE: \texttt{-{}-min <rmin> -{}-max <rmax>}

			where \texttt{rmin} and \texttt{rmax} are lower and upper bounds on the rewards occurring in the model, required by the mean payoff PE approach.
			\item CE: \texttt{-{}-global}
		\end{itemize}

		Note that the PRISM language does not allow to specify mean payoff properties as of now, hence we need to manually specify it.
		\item Mean payoff from reachability: \texttt{mean-payoff <model> <properties file> -{}-reachability <property name>}

		PE by default, add \texttt{-{}-global} to switch to CE
		\item Core: \texttt{core <model>}
	\end{itemize}

	All executions have \texttt{JAVA\_OPTS="-Xmx8064m -Xss128m"}, informing the JVM of the memory limits.

	\bigskip
	\item[\petone]
	Version 1.0 of \pet, presented in \cite{DBLP:conf/atva/Meggendorfer22}.
	It is also available at:
	\begin{center}
		\url{https://gitlab.lrz.de/i7/partial-exploration}
	\end{center}
	We use commit \texttt{840f269a34eb0be89521d4c3e47d87b1db26d3b9}, used in the submission for \cite{DBLP:conf/atva/Meggendorfer22}.

	Invocations are largely the same as for \pettwo.
	In particular, models, properties, and precision are specified as for \pettwo.
	
	\begin{itemize}
		\item Reachability: \texttt{reachability <model> <property>}

		PE by default, add \texttt{-{}-global} to switch to CE

		\item Mean payoff: \texttt{mean-payoff <model> -{}-reward <reward name>}
		\begin{itemize}
			\item PE: \texttt{-{}-reward-min <rmin> -{}-reward-max <rmax>}

			where \texttt{rmin} and \texttt{rmax} are lower and upper bounds on the rewards occurring in the model, required by the mean payoff PE approach.
			\item CE: \texttt{-{}-global}
		\end{itemize}
		\item Core: \texttt{core <model>}
	\end{itemize}
	We set the same environment variables as for \pettwo.

	\bigskip
	\item[\prismgames]
	Presented in \cite{prism-games} and available at:
	\begin{center}
		\url{https://github.com/prismmodelchecker/prism-games/}
	\end{center}
	We use release \texttt{v3.2} (the most recent version, \texttt{v3.2.1}, was only released days before the submission deadline).

	The used invocation is \texttt{<model file> <properties file> -{}-const <constants> -{}-property <property> -{}-epsilon 1e-6 -{}-intervaliter -maxiters 1000000 -javamaxmem 8064m -cuddmaxmem 4096m -javastack 128}.

	When using the explicit or hybrid engine, we append \texttt{-{}-explicit} or \texttt{-ddextraactionvars 64}, respectively.
	The latter is needed to circumvent errors for several models.
	Note that the hybrid engine does not support SGs.

	\bigskip
	\item[\prismext]
	Presented in \cite{EKKW22} and available at:
	\begin{center}
		\url{https://github.com/ga67vib/Algorithms-For-Stochastic-Games}
	\end{center}
	We used commit \texttt{e6e2e52af10e438b0d8da41cf7f44ebb910feb17} (the most recent at the time of writing).

	The invocation is as for \prismgames, only that we always use \texttt{-{}-explicit} engine and set \texttt{-maxiters 10}.
	The \texttt{maxiters} setting in \prismext has been repurposed to indicate how many VI steps should be performed before applying deflating.
	In \cite{EKKW22}, the authors recommended 10 as a good trade-off.

	\bigskip
	\item[\storm]
	Presented in \cite{storm} and available at:
	\begin{center}
		\url{https://github.com/moves-rwth/storm}
	\end{center}
	We used version \texttt{v1.8.1} (the most recent at the time of writing).

	The used invocation is \texttt{-{}-io:prism <model file> -pc -{}-io:constants <constants> -{}-io:prop <properties file> <property> -{}-sound -{}-precision 1e-6 -{}-absolute}.
	Communication with the authors of Strom confirmed that \texttt{-{}-sound} chooses the best available sound approach (in our version always topological optimistic value iteration).
	For mean payoff, instead of \texttt{-{}-io:prop ...}, we specify \texttt{-{}-io:prop 'R"<reward>"max=? [ LRA ]'}.

	For CE, we evaluated the (default) \texttt{sparse} and the \texttt{hybrid} engine.
	The hybrid engine uses Sylvan \cite{DBLP:journals/sttt/DijkP17} as its BDD library by default.
	Sylvan is optimized for parallelism and, additionally, Storm (apparently) does not detect on its own that it is restricted to a single CPU inside Docker, leading to massive performance penalties; concretely, we observed slowdowns by a factor of over 30.

	As recommended in communication with the authors, we add \texttt{-{}-ddlib cudd} to switch the BDD library back-end to CUDD, which is not parallelized (and thus does not need to pay the overhead of concurrent data structures).
	We also investigated adding the switch \texttt{-{}-sylvan:threads 1} to explicitly force Sylvan to use only one thread, however results were largely similar to using CUDD.

	For PE, we specify \texttt{-{}-core:engine expl}.
	Note that this version of Storm logs an incompatibility warning when using the \texttt{expl} engine, however communication with the authors confirmed that this warning can be ignored and will be fixed in future releases.

	\bigskip
	\item[\tempest]
	Presented in \cite{tempest} and available at:
	\begin{center}
		\url{https://github.com/tempest-shields/tempest-shields}
	\end{center}
	We used commit \texttt{58ec2ca3205cb375fcb085dc44687a3045e2f9cc} (the most recent at the time of writing).

	The used invocation is \texttt{-{}-io:prism <model file> -pc -{}-io:constants <constants> -{}-sound -{}-lra:nondetmethod vi -{}-lra:detmethod vi -{}-precision 1e-6 -{}-absolute -{}-prop <property expression>}, where \texttt{<property expression>} is the full PRISM property expression.
\end{description}

\subsection{Models}\label{app:exp-benchmarks}

\noindent Here we describe how and why we selected our benchmarks and give an overview of the landscape of probabilistic verification benchmarks.
Recall that we filter out instances of QVBS that are very simple or very time-consuming, see~\Cref{sec:exp-setup}.

\paragraph{MCs and MDPs with Reachability.}
Firstly, to provide a \enquote{standardized} comparison, we consider the quantitative verification benchmark set~\cite{qvbs} (QVBS).
Here, we select all (discrete- and continuos-time) Markov chains and MDPs given in the \prism modelling language with reachability objective.
(The QVBS does not include mean payoff.)
As the structure of a model dictates the performance of solution algorithms (see the final paragraph of \Cref{app:exp-add} for more details on this), we also included the model \model{mer}~\cite{mer}, which has been shown to be one of those that are interesting for the comparison of CE and PE algorithms~\cite{atva}.

\paragraph{MCs and MDPs with Mean Payoff.}
We consider all MDPs with mean payoff objectives that are listed in~\cite[App. G.1]{agarwal22}.
Most of these resulted from taking QVBS-models and complementing them with rewards (e.g.\ \model{ij}, \model{wlan}, \model{zeroconf}) or handcrafting simple examples (e.g.\ \model{counter}).
Several of these models fall into the \enquote{very simple} category and are thus removed from our benchmark set.
Moreover, several of them were transformed from reachability by obtaining a reward of 1 in the absorbing goal states. In essence, we transformed all SGs with reachability to such models by using our \enquote{reachability by mean payoff} approach in \pettwo, and thus exclude these as well.

Additionally, we also complemented several other models with rewards ourselves, e.g.\ \model{mer} and \model{phil-nofair}.

\paragraph{SGs with Reachability.}
We considered all models (i) that are on the \prismgames website\footnote{\url{http://www.prismmodelchecker.org/games/casestudies.php}}, (ii)~that were included at some point in the \prismgames GitHub\footnote{\url{https://github.com/prismmodelchecker/prism-games/tree/master/prism-examples/smgs}}, (iii)~that were included in the QComp 2023 comparison of tools for SGs~\cite{QComp23}, or (iv)~were used in at least one of the following papers~\cite{neller2004optimal,cloud,lex-journal,gandalf}.
We deliberately excluded the randomly generated models from~\cite{AEKSW22} since, due to their structure, a vast majority of the \enquote{solving} time is just spent in actually parsing the model.
Concretely, the models are specified in a monolithic fashion, while the PRISM modelling language is tailored towards composition.

Aside from the above, we also added variations of the \emph{pig game}~\cite{neller2004optimal}:
This dice-based game has been the topic of numerous papers (see \cite{piglinks} for a summary), since it is simple to describe, but difficult to solve.
Our results match those computed in the literature, validating the soundness of \pet.
Since this model has many probabilistic transitions but no non-trivial end components, this model \enquote{only} requires (lots of) Bellman updates, and thus particularly stress-tests this part of the implementation.


\paragraph{SGs with Mean Payoff.}
We considered the SGs that were created for \tempest in~\cite{tempest}, available at \url{https://tempest-synthesis.org/examples/}.
Note that the model \model{robotics\_optimal\_controller} cannot be parsed by \prismgames as it contains negative rewards.
Similarly, \model{robotics\_safety\_shield} contains no reward structures.
For \model{robotics\_path\_correction} and \model{traffic\_light\_shielding}, we adapted the property from synthesizing a shield to minimizing the mean payoff.
Notably, \tempest times out on both of these models, while \pettwo requires less than 20 and 2 seconds, respectively.

Moreover, we handcrafted models to test our implementation.
In this, we had two goals: testing soundness and scalability.
For soundness, we crafted several very small models to test the basic functionality of the algorithm.
We excluded them from the benchmark set due to their simplicity.
However, they are part of the artefact and can be used to further show the unsoundness and non-termination issues of \tempest.

For scalability, we created models that exhibit non-trivial structure, which we call \model{tree\_big\_mec}, \model{tree\_mul\_mec}, \model{tree\_mul\_sec}, \model{tree\_mul\_compl\_mec}, and \model{tree\_mul\_compl\_sec}.
All of these have the underlying structure of a full binary tree, but differ in what the nodes are (single states or simple or complex MECs or SECs).
They are scalable by a parameter $N$ which is the number of layers in the underlying tree.
This way, for small $N$ we can still validate the result, and for large $N$ (e.g.\ $N=20$, resulting in over a million nodes in the underlying tree), these models test the SEC-search and de-/inflating, with extremely large numbers of MECs or SECs, or a single extremely large MEC.
Note that, by the way they were constructed, they do not lend themselves to partial exploration, as they have to be fully explored.

\subsection{Further Details on the Evaluation}\label{app:exp-add}

\renewcommand{\plotmarksize}{2.5pt}
\begin{figure}[p]
\begin{minipage}{0.5\textwidth}
\begin{tikzpicture}
\begin{axis}[
	width=\textwidth,height=\textwidth,
	table/col sep=comma,
	x label style={anchor=north,inner sep=0pt},
	y label style={anchor=south,inner sep=0pt},
	xtick={2,10,60}, xticklabels={2,10,60},
	ytick={2,10,60}, yticklabels={2,10,60},
	xmin=1,ymin=1,xmax=\legendmax,ymax=\legendmax, xmode=log,ymode=log,
	axis x line*=bottom,
	axis y line*=left,
	xlabel=\texttt{\petone CE},ylabel=\texttt{\pettwo CE},
]
	\axislines

	\pettable table [x=pet_1_g, y=pet_2_g] {exp-results/current/csv_pair_1_vs_2_ce.csv};
\end{axis}
\end{tikzpicture}%
\end{minipage}%
\begin{minipage}{0.5\textwidth}
\begin{tikzpicture}
\begin{axis}[
	width=\textwidth,height=\textwidth,
	table/col sep=comma,
	x label style={anchor=north,inner sep=0pt},
	y label style={anchor=south,inner sep=0pt},
	xtick={2,10,60}, xticklabels={2,10,60},
	ytick={2,10,60}, yticklabels={2,10,60},
	xmin=1,ymin=1,xmax=\legendmax,ymax=\legendmax, xmode=log,ymode=log,
	axis x line*=bottom,
	axis y line*=left,
	xlabel=\texttt{\pettwo PE},ylabel=\texttt{\pettwo CE},
]
	\axislines
	\pettable table [x=pet_2_s, y=pet_2_g] {exp-results/current/csv_pair_pe_vs_ce.csv};
\end{axis}
\end{tikzpicture}%
\end{minipage}%

\bigskip

\begin{minipage}{0.5\textwidth}
\begin{tikzpicture}
\begin{axis}[
	width=\textwidth,height=\textwidth,
	table/col sep=comma,
	x label style={anchor=north,inner sep=0pt},
	y label style={anchor=south,inner sep=0pt},
	xtick={2,10,60}, xticklabels={2,10,60},
	ytick={2,10,60}, yticklabels={2,10,60},
	xmin=1,ymin=1,xmax=\legendmax,ymax=\legendmax, xmode=log,ymode=log,
	axis x line*=bottom,
	axis y line*=left,
	xlabel=\texttt{\storm PE},ylabel=\texttt{\pettwo PE},
]
	\axislines

	\stormtable table [x=Storm-explore, y=pet_2_s] {exp-results/current/csv_pair_pe_vs_storm.csv};
\end{axis}
\end{tikzpicture}%
\end{minipage}%
\begin{minipage}{0.5\textwidth}
\begin{tikzpicture}
\begin{axis}[
	width=\textwidth,height=\textwidth,
	table/col sep=comma,
	x label style={anchor=north,inner sep=0pt},
	y label style={anchor=south,inner sep=0pt},
	xtick={2,10,60}, xticklabels={2,10,60},
	ytick={2,10,60}, yticklabels={2,10,60},
	xmin=1,ymin=1,xmax=\legendmax,ymax=\legendmax, xmode=log,ymode=log,
	axis x line*=bottom,
	axis y line*=left,
	xlabel=\texttt{\prismgames{} / \storm (hybrid)},ylabel=\texttt{\pettwo CE},
]
	\axislines

	\prismtable table [x=PRISMh, y=pet_2_g] {exp-results/current/csv_pair_ce_vs_prism_h.csv};
	\stormtable table [x=Storm-hybrid, y=pet_2_g] {exp-results/current/csv_pair_ce_vs_storm_h.csv};
\end{axis}
\end{tikzpicture}%
\end{minipage}%

\bigskip

\begin{minipage}{0.5\textwidth}
\begin{tikzpicture}
\begin{axis}[
	width=\textwidth,height=\textwidth,
	table/col sep=comma,
	x label style={anchor=north,inner sep=0pt},
	y label style={anchor=south,inner sep=0pt},
	xtick={2,10,60}, xticklabels={2,10,60},
	ytick={2,10,60}, yticklabels={2,10,60},
	xmin=1,ymin=1,xmax=\legendmax,ymax=\legendmax, xmode=log,ymode=log,
	axis x line*=bottom,
	axis y line*=left,
	xlabel=\texttt{\prismgames},ylabel=\texttt{\pettwo CE (reach as MP)},
]
	\axislines

	\prismtable table [x=PRISMe, y=pet_2_g_rmp] {exp-results/current/csv_pair_ce_rmp_vs_prism.csv};
\end{axis}
\end{tikzpicture}%
\end{minipage}%
\begin{minipage}{0.5\textwidth}
\begin{tikzpicture}
\begin{axis}[
	width=\textwidth,height=\textwidth,
	table/col sep=comma,
	x label style={anchor=north,inner sep=0pt},
	y label style={anchor=south,inner sep=0pt},
	xtick={2,10,60}, xticklabels={2,10,60},
	ytick={2,10,60}, yticklabels={2,10,60},
	xmin=1,ymin=1,xmax=\legendmax,ymax=\legendmax, xmode=log,ymode=log,
	axis x line*=bottom,
	axis y line*=left,
	xlabel=\texttt{\pettwo CE (reach as MP)},ylabel=\texttt{\pettwo CE},
]
	\axislines

	\pettable table [x=pet_2_g_rmp, y=pet_2_g] {exp-results/current/csv_pair_ce_vs_ce_rmp.csv};
\end{axis}
\end{tikzpicture}%
\end{minipage}%

\caption{
		Further comparisons of \pettwo to other tools.
		From left to right, top to bottom, we compare {\pettwo}-CE and {\petone}-CE, {\pettwo}-CE and {\pettwo}-PE, {\pettwo}-PE and \storm-PE, \pettwo-CE and \prismgames (\markincaption{Dark2-C}{x}) / \storm (\markincaption{Dark2-B}{star}) with \emph{hybrid} engine, \pettwo-CE with mean payoff from reachability and \prismgames, and \pettwo-CE with mean payoff from reachability with \pettwo-CE.
		We use the same graph presentation as in \cref{fig:results}.
}
\label{fig:app-results}
\end{figure}
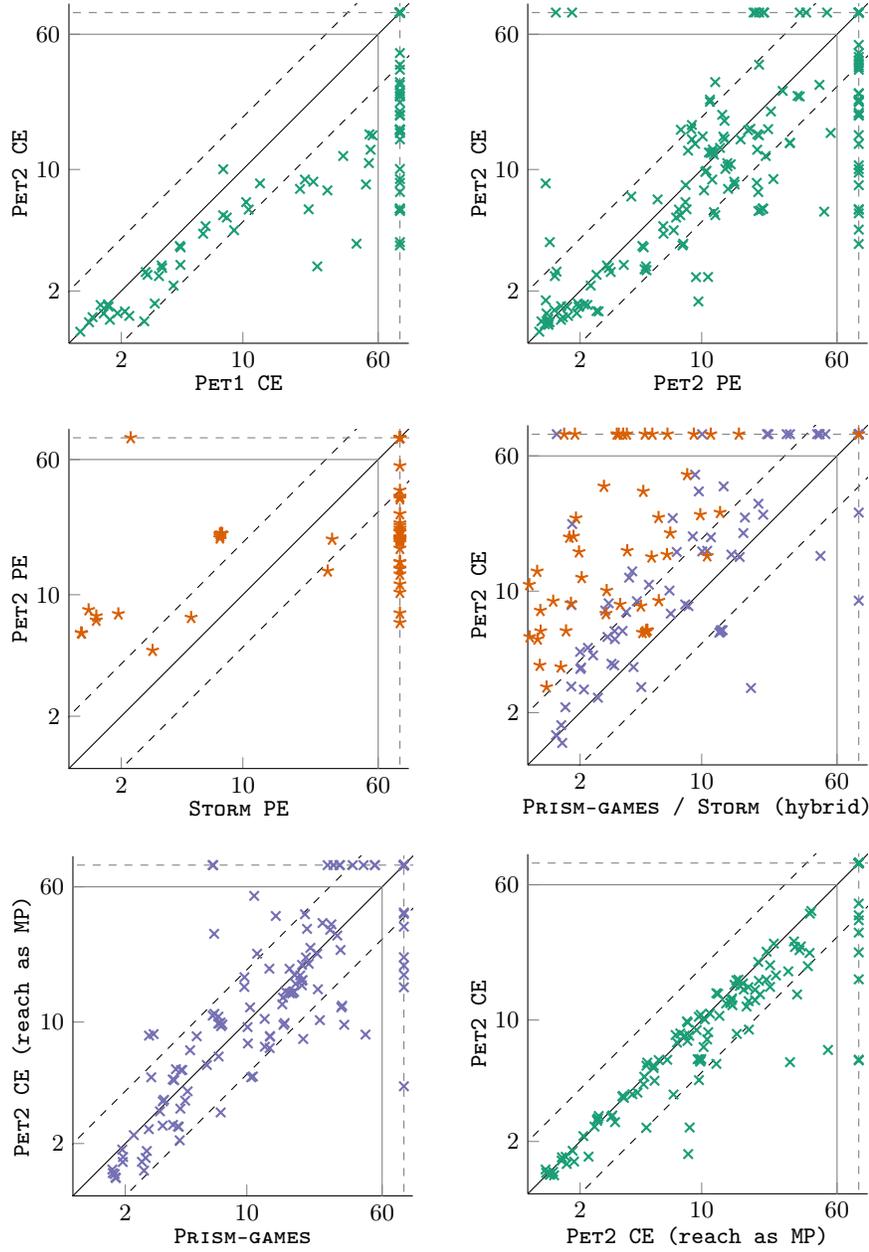

\paragraph{Additional Results.}
We present additional comparisons in \cref{fig:app-results} and briefly discuss the findings.

Comparing the CE implementation of \pettwo and \petone, we see drastic improvements.
This is mainly due to \petone implementing a \enquote{fake} CE by pretending to the PE algorithm that in every step all states of the system are sampled.
This comes (i)~with additional overhead of fitting the whole system into the \enquote{sampled states} data structure and (ii)~prohibits the use of CE-specific optimizations (such as only searching for ECs once).

For CE vs.\ PE, we see that there are extremes in both directions, again underlining that it very much depends on the model structure which of the two approaches is faster, and quality of implementation is secondary.

Comparing the PE approaches of \pettwo and \storm, we see that when \storm terminates, it usually is quite fast, however it fails to terminate on numerous models.
We believe that this is due to the PE implementation of \storm is missing a few optimization heuristics present in \pettwo.

Next, comparing to the hybrid engines of \prismgames and \storm makes for a slightly unfair comparison, since the approaches are different.
Moreover, the hybrid engine does not support SGs, so this comparison is restricted to MCs and MDPs with reachability.
In a nutshell, the hybrid approach stores parts of the system symbolically and parts explicitly, sometimes allowing for exponential gains due to symbolic representation.
Overall, \prismgames with hybrid engine seems to be faster (which also shows in a score of \cmpscore{0.7}{55}{2}{10}), but, similar to comparing CE and PE, there are extremes in both directions.
Again, as expected, Storm outperforms its competition (\cmpscore{0.2}{53}{0}{13}).

Finally, we also discuss the comparison of the \enquote{reachability by mean payoff}-approach to \prismgames and the dedicated approach of \pettwo.
For \prismgames, we see that performance is overall on par, with a noticeable amount of variance.
For \pettwo, interestingly, on many models the dedicated approach performs nearly identical.
We conclude that these are models that have barely any internal structure that can be exploited by graph analysis; to get a result we simply need to apply Bellman updates.
For future improvements and, in particular, evaluating reachability-specific improvements to, e.g., iteration order, these models might be good candidates to test for the effects of such optimizations.


\paragraph{Soundness.}

We summarize all wrong results we observed during our evaluation.
\begin{itemize}
	\item
	\pettwo and \storm had no wrong results.
	\item
	\petone gave 5 wrong answers, all of which were related to two peculiar bug in (i)~the aperiodicity transformation that is necessary for solving models with mean payoff objectives and (ii)~handling of special cases for distributions.
	Both of these bugs have been discovered and fixed in \pettwo.
	\item
	\prismgames gave 6 wrong answers, all of which seem to be due to \prismgames using the classical but unsound stopping criterion of value iteration, i.e.\ terminating once the iterates hardly change.
	In practice, this often is a good indicator that the iterates are reasonably close to the true result, yet does not give any formal guarantees.
	This reflects in the magnitude of the errors:
	While the results were off by more than the allowed precision, it was only by a small amount (never more than $2\cdot \varepsilon$).
	On the one hand, this shows that the extreme differences exhibited by handcrafted models such as the one in~\cite[Fig. 3]{HM18} do seldom occur in practice, but, on the other hand, the underlying problem indeed occurs even in realistic models not chosen antagonistically for naive VI.
	\item \tempest gave 6 wrong answers, often by a significant amount.
	In particular, the one reported in the main body (0.0003 instead of 0.48) happened on the model \model{pigs} with a reachability objective.
	Thus, there seem to be further problems than only the naive stopping criterion of value iteration, especially due to the rather \enquote{simple} structure of \model{pigs} we explained in \Cref{app:exp-benchmarks}.
	
	We mention that for mean payoff on SGs, \tempest applies the stopping criterion of~\cite{DBLP:conf/cav/AshokCDKM17}, which is only proven to work for MDPs.
	Applying this criterion is also sound on SGs, however incomplete, as it is missing the detailed MEC analysis techniques of~\cite{lics23} and may lead to non-termination.
	We verified this using several small handcrafted examples (included in the artefact).
	Surprisingly, we observed both non-terminating behaviour as well as termination with wrong results, depending on whether relative or absolute error is considered.
	In particular, the former leads to wrong results, while the latter leads to non-termination.
	This suggests that many of the timeouts of \tempest would also result in wrong answers when using the relative criterion.
\end{itemize}
We conclude with noting that most of the models we used for this evaluation do not have reference results, especially for stochastic games.
Our complete benchmark collection (found in the artefact) provides several small games with reference results (on which \tempest often provides wrong answers or times out), however these were excluded due to their miniscule size (mostly less than 10 states).

\paragraph{Details on Model Structure.}
Reiterating~\cite{AEKSW22,DBLP:conf/atva/Meggendorfer22,DBLP:conf/tacas/HartmannsJQW23}, we again underline that the structure of a model is the leading factor for the (relative) performance of algorithmic approaches.
This is supported by our comparison of CE vs.\ PE vs.\ hybrid approaches above.

In particular, the size of a model only loosely correlates with the solution time:
\pettwo can solve non-trivial models with millions of states within a minute, e.g.\ \model{zeroconf} or \model{tree\_mul\_compl\_mec}, the latter of which requires non-trivial search for SEC-candidates and de-/inflating.
Moreover, the former can be artificially inflated to arbitrary sizes while PE approaches take constant time, see e.g.\ \cite{DBLP:conf/atva/Meggendorfer22}.
In contrast, the adversarial example in~\cite[Fig. 3]{HM18} (with e.g.\ $n = 100$) is a Markov chain with less than hundred states, a branching factor of two, i.e.\ only a few hundred transitions, only one non-trivial SCC, and a \emph{minimal} transition probability of $\frac{1}{2}$.
In other words, this model is \enquote{as simple as can be} from its graph-theoretic properties, yet all VI-based approaches time out on this adversarial example, since it actually requires an exponential number (in $n$) of Bellman updates.
Moreover, by pushing up $n$ a bit further, the propagated changes are so small that standard  IEEE 754 64 bit \texttt{double} precision is insufficient.

We observed that, as a general theme, most of the models provided by the benchmark suites we referenced have a surprisingly simple structure, often not even including non-trivial end components.
For example, a significant number of reachability instances (43 of the total 114 considered) are solved by \pettwo's graph analysis, not requiring any Bellman updates at all.
Thus, a lot of the performance boils down to (i)~how fast can the model be parsed and generated and (ii)~how quickly can graph analysis and the Bellman operator be applied.
As such, the established model sets are only of limited use to evaluate the performance improvement gained by more sophisticated optimizations.
Rather, their simplicity often prohibits the application of such more complex optimizations and may even suggest a decrease in performance when such optimizations incur an overhead.
This, for example, applies for \pettwo with predecessor tracking:
There are structurally simple models where we cannot gain anything from graph analysis, thus the predecessor sets are built in vain, unnecessarily costing space and time.
However, this optimization itself pays off a lot especially on complex models.

Regarding the graph analysis, we note that for two models that \pettwo solves by graph pre-computations, \prismgames has to resort to Bellman updates.
Upon further investigation, we found that this is due to \pettwo treating states with only one choice specially during graph analysis and could replicate the difference on a four state model.

}{}

\end{document}